\documentclass[fleqn,usenatbib]{mnras}

\usepackage[T1]{fontenc}

\DeclareRobustCommand{\VAN}[3]{#2}
\let\VANthebibliography\thebibliography
\def\thebibliography{\DeclareRobustCommand{\VAN}[3]{##3}\VANthebibliography}

\usepackage{graphicx}

\usepackage{amsmath}

\usepackage{amssymb}
\usepackage[dvipsnames]{xcolor}

\def\be{\begin{equation}}
\def\ee{\end{equation}}
\def\ba{\begin{equation}\begin{array}}
\def\ea{\end{array}\end{equation}}
\def\bi{\begin{itemize}}
\def\ei{\end{itemize}}
\def\mb{\mathbf}

\title[FDM halos with self-interactions]{Fuzzy dark matter halos with  repulsive self-interactions: coherent soliton and halo vortex network with moderate self-coupling}

\author[M. Indjin {\it{et al.}}]{
Milos Indjin, 
Nick Keepfer, 
I-Kang Liu\thanks{E-mail: i-kang.liu@newcastle.ac.uk},
Nick P. Proukakis\thanks{E-mail: nikolaos.proukakis@newcastle.ac.uk},
and Gerasimos Rigopoulos\thanks{E-mail: gerasimos.rigopoulos@newcastle.ac.uk}
\\
School of Mathematics, Statistics and Physics,\\
Newcastle University, Newcastle upon Tyne, NE1 7RU, United Kingdom\\
}

\date{Accepted XXX. Received YYY; in original form ZZZ}

\pubyear{2025}

\begin{document}
\label{firstpage}
\pagerange{\pageref{firstpage}--\pageref{lastpage}}
\maketitle

\begin{abstract}
We examine the impact of moderate repulsive self-interactions on fuzzy dark matter halos generated by merging smaller Gaussian density concentrations. We study the size of the core and the granules, the spatial dependence of the field's coherence, the turbulent vortex tangle and the oscillation frequency of the central soliton, covering the range from quantum-pressure-dominated to self-interaction-dominated stabilisation of the solitonic core. For the probed self-coupling strengths $g$ and with a fixed initial configuration, mergers with increasing $g$ result in cores with increased size and a reduced central density, oscillating with decreased frequency, in accordance with expectations from the study of isolated Self-interacting Fuzzy Dark Matter (SFDM) solitons. By contrast, the characteristic granule size and typical inter-vortex distances in the surrounding halo are only mildly affected, growing much less relative to the core. The total length of the vortex network, although less robust, shows no signs of decay over our simulation timescales. The generated halos therefore develop central  self-interaction-dominated cores, but with the outer halos still supported by quantum-pressure and classical kinetic energy in equipartition as in the non-interacting case.
Furthermore, measures of coherence of the field clearly separate the condensed core, identified via the Penrose-Onsager (largest eigenvalue) mode of the entire classical field, from the surrounding quasi-coherent halo.
Unlike the $g=0$ case, we observe a relative increase of incoherent fluctuations coexisting with the coherent mode at the centre of the halo with increasing $g$, a phenomenon also observed in laboratory condensates at non-zero temperature. 

\end{abstract}




\section{Introduction}

The Fuzzy Dark Matter (FDM) model has been receiving increasing attention in the literature as an attractive alternative to Cold Dark Matter (CDM)
(\cite{Schive2014,Schive2014a,Marsh2015, 2016PhR...643....1M,Mocz2017,Bernal2017,
Lin2018,2018MNRAS.478.2686C,Veltmaat2018,Levkov2018,
Mocz_2019, Hui2021,2021A&ARv..29....7F,Matos_Review_2024,Berezhiani:2025maf,May2021StructureDynamics,Yavetz2021,Chan2022,Dome2023,
Liu2023,OHare2024,schive2025}), with scattered original studies of such scalar dark matter already stretching back decades - see e.g.~\citep{Baldeschi:1983mq, 1985MNRAS.215..575K, Membrado:1989bqo,PhysRevD.50.3650, Lee:1995af,Hu2000, Bohmer2007}.
In recent years, the basic parameter of the model that has mostly been considered is the mass of the boson particle, normally assumed to be within (or close to) the range $m \sim \mathcal{O}(10^{-22} - 10^{-20}) \, eV/c^2$ for which wave effects are relevant on galactic scales.

\begin{figure*}

	\includegraphics[width=1.95\columnwidth]{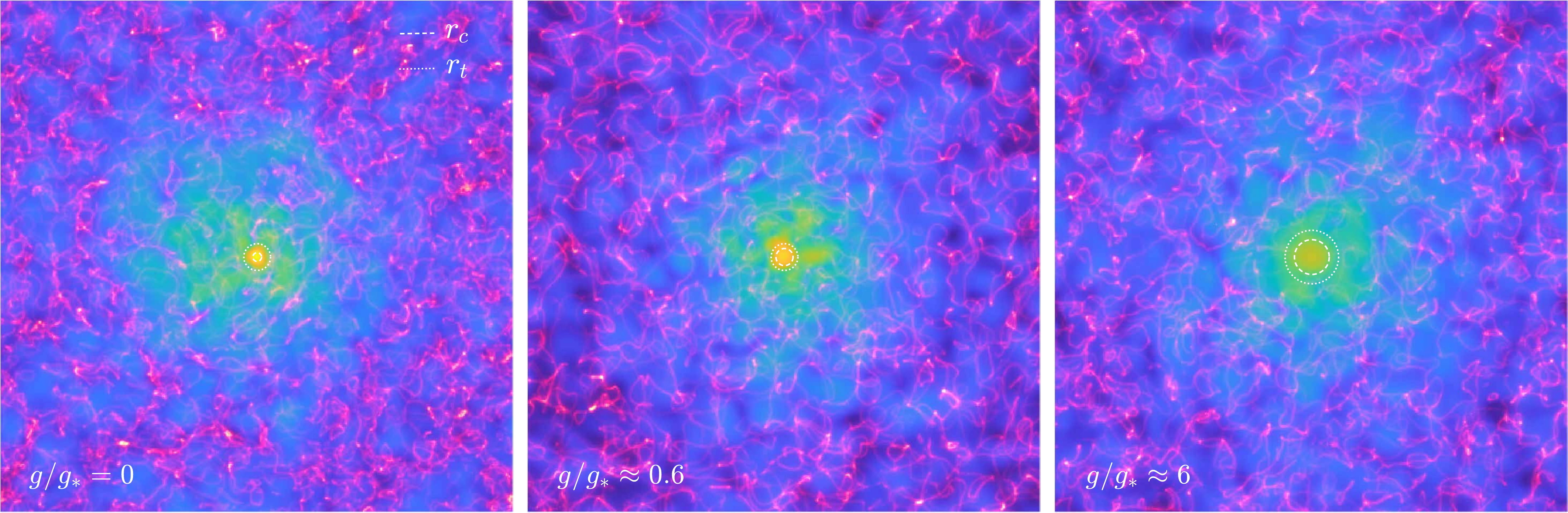}
    \caption{Visualisation of the dependence of the soliton core and surrounding vortex-infilled turbulent halo on the repulsive boson self-coupling $g$.
    Shown are volume rendering images of both density and vortices for (left) a non-self-interacting FDM halo ($g/g_\ast=0$) and self-interacting ones with (middle) ($g/g_\ast\approx0.6$) and (right) ($g/g_\ast\approx 6$), where
    $g_\ast$ [defined in Eq.~(\ref{eq:g_*})] is a characteristic self-interaction strength determined from the solitonic core that demarcates the transition from weak to strong self-interaction. 
    The white dashed and dotted lines mark the core radius $r_c$ and transition radius $r_t$ respectively, extracted from the core-halo fit for the radial density profiles. As the self-interaction strength increases from left to right, the core density decreases but there is no significant influence on the granule size in the surrounding halo. This is because the density in the outer halo is too low for the self-interaction to have a  significant impact there. All halos are formed from the same initial spatial configuration of 10 randomly distributed Gaussians with slightly different masses and widths set by the $g=0$ virial theorem.}
    \label{fig:1}
\end{figure*}

The inclusion of local self-interactions, often but not exclusively assumed to be of quartic type and of strength $g \neq 0$, can have important implications, both for the evolution of linearized density perturbations from the early universe but also for non-linear halos and the characteristic solitonic objects that form in their centres -- see e.g.~\citep{Harko2011zt, Chavanis2011, Chavanis2011Delfini,Rindler-Daller:2011afd, Rindler-Daller2014, Li2014, Fan:2016rda,Li2017, Desjacques:2017fmf, Dawoodbhoy2021, Shapiro2021, Glennon2020-2, Hartman2022, Chakrabarti2022, chanda_formation, marsh_formation, Foidl2023, Indjin2024, Valageas2024, RindlerDaller2025,Sivakumar2025,
Mocz2023, Mocz2024, Moss:2024mkc} for a non-comprehensive reference list where self-interactions are included. It is therefore natural to examine the detailed effects of such a self-interacting fuzzy dark matter, henceforth termed as SFDM \citep{RindlerDaller2025}, to the characteristic core-halo structure emerging in gravitationally bound FDM halos; this is particularly relevant and timely, given the recent evidence that non-zero ($g \neq 0$) self-interactions appear essential in order to describe galactic rotation curves with a {\em single} value of the boson mass~(\cite{Delgado2022,Indjin:2025lqs}), thereby resolving an identified earlier problem encountered with non-interacting FDM (see, \textit{e.g.}~\cite{Bernal2017,Meinert2021,Banares-Hernandez2023b,Khelashvili2023} and references therein).



In a previous work~(\cite{Liu2023}), we considered the density and coherence properties of a single virialised FDM cored-halo in the absence of a boson self-interaction. The cored-halo was formed from the gravitational coalescence of ten idealized solitons of comparable masses, randomly distributed around the central point of our numerical grid under the constraint of a centrally-located centre of mass, and a prescribed total mass. As expected, the spatial profile of such a non-interacting cored-halo  was found to be well fit by the combination of a centrally-located empirical solitonic profile, combined with an NFW profile in the outer regions~\citep{Schive2014,Schive2014a,Marsh2015,Mocz2017,Chan2022}. 

Performing a detailed analysis of the spatial coherence via the commonly-used Schr\"{o}dinger-Poisson equations (SPE), \citep{Liu2023} demonstrated that such a bimodal spatial configuration involves a central, {\em fully coherent} Bose–Einstein
condensate (BEC), 
embedded within a turbulent `quasi-condensate' halo.
The central soliton was identified with a pure BEC, through numerical calculation of the mode of the single-particle density matrix with the largest eigenvalue, known as the Penrose-Onsager (PO) mode~\citep{1956PhRv..104..576P,2008AdPhy..57..363B}. This coherent solitonic core was confirmed as arising via a stabilization between gravitational attraction and the characteristic quantum pressure of FDM, revealing also evidence of dynamical core oscillations found to be anti-correlated to the peak location of the power spectrum of the whole halo. 

It is important to point out that, depending on how the system is initialized, solving for the entire classical field of the fuzzy dark matter model can in general include both the "coherent" condensate degrees of freedom as well as the "incoherent" modes above the condensate:
This important point, which is key to our ability to probe both condensate and incoherent dynamics within a single dynamical classical field simulation, will be further discussed in Sec.~\ref{sec:coherence-gppe}.

The solitonic core is distinct from the surrounding outer halo with a clear crossover region between them. The halo surrounding the coherent core was found to behave as incoherent on average, while exhibiting local semi-coherent patches both in space and time, arising from phase and density fluctuations; such patches correspond to the structures known as ``granules'' in the literature. These semi-coherent patches were intertwined with a slowly evolving, but seemingly non-decaying, tangle of quantum vortices\footnote{See also related earlier work on FDM turbulence by~\cite{Woo_2009,Mocz2017}.}, exhibiting sustained quantum, or `Vinen', turbulence~\citep{Baggaley2012} rather than the better known Kolmogorov turbulence. Specifically, an analysis of the incompressible kinetic energy spectrum revealed a characteristic peak marking the inter-vortex distance, and a $k^{-3}$ scaling at larger $k$. The characteristic peak of such spectrum, was found to essentially coincide with the peak of the granule power spectrum characterizing the typical granule size.

It is natural to ask in what manner, and to what extent, the above features of non-interacting FDM halos would hold, or be critically modified through self-interactions. The aim of the present work is to extend our earlier analysis of~\citep{Liu2023} in the context of SFDM, focusing here on repulsive interactions of moderate strength, as will be quantified below. 

We perform a number of merger simulations with identical initial spatial configurations but increasing self-coupling strength $g$. The latter is primarily found to lead to: i) a reduced central density~(\cite{Chavanis2011,Indjin2024}), ii) a flattening out of the enlarged solitonic core which becomes increasingly stabilised by repulsive interactions with quantum pressure becoming less relevant and iii) an associated slight reduction in the core's central coherence due to condensate depletion. Outside the solitonic core, interactions are found to have much less of an effect, and the (quantum) turbulent structure consistent with the density granules remains dynamically sustained, at least for the interaction strengths studied here. This is consistent with the picture put forward in \citep{Dawoodbhoy2021} where a self-interaction-supported core in the Thomas-Fermi regime is surrounded by a halo where self-interactions do not play a prominent role and it is the the quantum pressure that still sets the local de Broglie wavelength as the characteristic scale. The transition from non-interacting to interaction-dominated in our simulations can be seen in Fig.~\ref{fig:1}. We further find that the
size of the central core and its lowest radial oscillation frequency agree quite well with numerical predictions for isolated self-interacting solitons \citep{Indjin2024}. The core's shape for $g > 0$ can be very well approximated by a super-Gaussian profile \citep{Indjin:2025lqs} which can be used to accurately extract the core from the surrounding halo.   

This paper is structured as follows: 
Sec.~\ref{sec:gppe} gives an overview of the SFDM model, in the form of coupled Gross-Pitaevskii-Poisson equations (GPPE), with emphasis on the different contributions to the energy functional. In Sec.~\ref{sec:analytical} 
we identify a characteristic self-interaction strength $\Gamma_g$ demarcating the transition from non-interacting to moderately interacting solitons, and recall pertinent points from the analysis of \citep{Indjin2024, Indjin:2025lqs} giving semi-analytical estimates for the interaction-strength-dependent peak density and the core radius, as well as for a characteristic radial oscillation frequency. We also provide details on our super-Gaussian fit for the SFDM solitonic density profiles. In Sec.~\ref{sec:Numerics} we present our numerical procedure and the merger simulations. Sec.~\ref{sec:profiles} discusses our numerical results for the effect of weak to moderate interactions on the resulting equilibrium cored-halo density profiles and their coherence, clarifying the manner in which the numerical solution of the GPPE encompasses both "coherent" (condensate) and "incoherent" parts of the system and commenting on the observed central condensate depletion. We also compare theoretical predictions for the central density and its oscillation frequency from isolated SFDM solitons to those measured in the halo cores of our simulations, finding good agreement. Sec.~\ref{sec:energy} discusses the impact of self interactions on relevant power spectra and vortex network characterization, showing that the outer halos are dominated by quantum pressure and classical kinetic energy in equipartition, making them similar to their $g=0$ counterpart. Our findings are summarized and discussed in Sec.~\ref{sec:conclusions}.

\section{Gross-Pitaevskii-Poisson equations} \label{sec:gppe}

\subsection{Self-Interacting FDM}

In the presence of interactions, the SFDM field evolves according to the coupled GPPE~\citep{Bohmer2007, Chavanis2011, Rindler-Daller:2011afd}
\be\label{eq:GPE1}
i\hbar\frac{\partial}{\partial t}\Psi(\mb{r},t)=\left[-\frac{\hbar^2\nabla_\mathbf{r}^2}{2m}+g\rho(\mb{r},t)+\Phi(\mb{r},t)\right]\Psi(\mb{r},t)
\ee
\be\label{eq:Poisson_0}
\nabla_\mathbf{r}^2\Phi(\mb{r},t)=4\pi Gm\left[
\rho(\mb{r},t)-\langle\rho(\mb{r},t)\rangle
\right].
\ee
The SFDM field evolution is characterized by a local self-interacting contribution $g \rho$, where $\rho(\mb{r},t)=|\Psi(\mb{r},t)|^2$ is the mass density, and a Newtonian particle gravitational potential energy $\Phi$.
Here $g$
is the  self-interaction strength 
and the average density $\langle\rho(\mb{r},t)\rangle$ has been subtracted in the Poisson equation to regularize the contribution from a possible non-zero average background (\cite{binney2008galactic})\footnote{Subtraction of the average density from the Poisson equation for the gravitational potential Eq.~\eqref{eq:Poisson_0}, often referred to as the `Jeans Swindle', arises in the self-consistent study of perturbations around infinite homogeneous gravitating systems within General Relativity. It also makes the solution satisfy the periodic boundary condition~\citep{Dabo2008}.}. 
We note here that the mass density is normalized according to
\be
M=\int d\mathbf{r}\left|\Psi(\mathbf{r,t})\right|^2,
\ee
which is conserved in time, with $(M/m)$ corresponding to the SFDM particle number.
In the limit of $g=0$, the GPPE reduces to the Schr\"{o}dinger-Poisson equation~(SPE), which has been widely studied in the literature 
(e.g.~\cite{Schive2014a,Marsh2015,Mocz2017}).
In this work, we specifically focus on the repulsive self interaction, $g\geq0$. 
Note that, as discussed in Sec.~\ref{sec:coherence-gppe}, unless explicitly generated to be in the ground state, the SFDM classical field $\Psi({\bf r},t)$ should not in general be identified as the "condensate" mode but may also contain "incoherent" features embedded within it.

A hydrodynamic description of SFDM in terms of the mass density $\rho$ and a velocity field $\mb{v}$ can be obtained by introducing the Madelung transformation to the wavefunction, $\Psi=\sqrt{\rho}e^{i\varphi}$, where the velocity is given by 
$\mb{v}(\mb{r},t)=(\hbar/m)\nabla\varphi(\mb{r},t)$.
One thus obtains a continuity equation for $\rho$ in the form

\be
\frac{\partial}{\partial t}\rho+\nabla\cdot(\rho\mb{v})=0
\ee
and a corresponding equation for the velocity $\mb{v}$
\be
\frac{\partial}{\partial t}\mb{v}+\frac{\nabla}{m}\left[
\frac{m|\mb{v}|^2}{2}+g\rho+\Phi-\frac{\hbar^2}{2m}\frac{\nabla^2\sqrt{\rho}}{\sqrt{\rho}}
\right]=0\,, \label{eq:v}
\ee
expressed in terms of self-interaction, gravitational potential and quantum pressure contributions. 
This equation has the form of a classical irrotational fluid equation in the absence of dissipation, with the only addition being the quantum pressure term which disappears in the limit $\hbar \rightarrow 0$. One might therefore expect the presence of classical turbulent features in FDM, with modifications emerging from the quantum pressure term which is primarily relevant in regions of high density gradients. 
Since the study of vortices in the context of such hydrodynamic equations requires specialised handling with further assumptions~(see, e.g.,~\cite{cfb-review}), it proves numerically convenient for studies of vortex dynamics to solve the GPPE instead of the above hydrodynamic equations.

\subsection{Energy Contributions}

One can define the energy functional of the system which receives three distinct contributions:
\be\label{eq:energy_functional}
E\left[\Psi^\ast,\Psi\right]=E_{\nabla^2}\left[\Psi^\ast,\Psi\right]+E_g\left[\Psi^\ast,\Psi\right]+E_\Phi\left[\Psi^\ast,\Psi\right]
\ee
where
\be
E_{\nabla^2}\left[\Psi^\ast,\Psi\right]=-\frac{1}{m}\int d\mathbf{r} \, \Psi^\ast\frac{\hbar^2\nabla^2}{2m}\Psi \,,
\ee
\be
E_g\left[\Psi^\ast,\Psi\right]=\frac{1}{2m}\int d\mb{r} \, g\left|\Psi\right|^4
\ee
and 
\be
E_\Phi\left[\Psi^\ast,\Psi\right]=\frac{1}{2m}\int d\mb{r} \, \Phi\left|\Psi\right|^2
\ee
are the quantum kinetic, self-coupling and gravitational potential energies, respectively. The gravitational potential $\Phi$ is the solution to \eqref{eq:Poisson_0} and can also be explicitly expressed as 
\begin{equation}
\Phi(\mb{r},t) = - \int d\mb{r}^\prime \frac{Gm}{|\mb{r}-\mb{r}'|} \left[
\rho(\mb{r}^\prime,t)-\langle\rho\rangle
\right].
\end{equation}
The Hartree variational principle, $i\hbar\partial_t\Psi=m\delta E/\delta\Psi^\ast$, then directly gives the GPPE.

The use of the Madelung transformation allows for the $E_{\nabla^2}$ term to be further split into two, allowing the energy of the system, Eq.~(\ref{eq:energy_functional}) to be rewritten as a sum of four different parts,
\ba{rl} \label{eq:energy}
E=&\displaystyle E_\mathrm{ke}+E_\mathrm{qp}+E_g+E_{\Phi}\\\\
=&\displaystyle \int d\mb{r}\left[\varepsilon_\mathrm{ke}(\mb{r})+\varepsilon_\mathrm{qp}(\mb{r})+\varepsilon_g(\mb{r})+\varepsilon_{\Phi}(\mb{r})\right]
\ea
where $E_\mathrm{ke}$, $E_\mathrm{qp}$, $E_g$ and $E_\Phi$ are classical kinetic, quantum pressure, self-interaction and gravitational potential energies respectively, with the energy densities contributed by the Laplacian term given by
\be
\varepsilon_\mathrm{ke}(\mb{r})=\frac{1}{2}\rho(\mb{r})|\mb{v}(\mb{r})|^2,\quad \varepsilon_\mathrm{qp}(\mb{r})=\frac{\hbar^2}{2m^2}\left|\nabla\sqrt{\rho(\mb{r})}\right|^2\,,
\ee
while 
\be
\varepsilon_g(\mb{r})=\frac{g}{2m}\left[\rho(\mb{r})\right]^2\,,
\ee
and
\be
\varepsilon_{\Phi}(\mb{r})=\frac{1}{2m}\Phi(\mb{r})\rho(\mb{r})\;.
\ee

In the absence of interactions, the formation of vortices and their existence in FDM halos was numerically demonstrated in~\citep{2017MNRAS.471.4559M, 2021JCAP...01..011H, Liu2023}.
While the velocity field $\mb{v}$ diverges at a vortex core, the current   
\be
\mb{F}=\sqrt{\rho}\mb{v}
\ee
tends to a constant at the centre of a vortex and plays a key role in the study of superfluid turbulence.
We decompose the velocity field into compressible (irrotational) and incompressible (rotational) parts via the Helmholtz decomposition, with
\be
\mb{v}=\mb{v}^c+\mb{v}^i \textrm{ and }\mb{F}=\mb{F}^c+\mb{F}^i
\ee
satisfying
\be
\nabla\times\mb{v}^c=\nabla\times\mb{F}^c=0\textrm{ and }\nabla\cdot\mb{v}^i=\nabla\cdot\mb{F}^i=0
\ee
both of which can be computed via Fourier transformation~\citep{2005JPSJ...74.3248K}.
The rotational component of the velocity field close to the vortex cores is associated with a velocity profile 
$\left|\mb{v}^i \right| \propto 1/r$,
where $r$ is the distance from the vortex centre, whose large value localized around the vortices can be used to probe the vortex positions.  
To probe the superfluid turbulent spectra~\citep{1997PhRvL..78.3896N, 1997PhFl....9.2644N, 2005JPSJ...74.3248K, 2010PhRvA..81f3630N},
the classical kinetic energy is also decomposed into compressible and incompressible parts via
\be
E_\mathrm{ke}=E_\mathrm{ke}^c+E_\mathrm{ke}^i
\ee
with the corresponding energy densities 
\be
\varepsilon_\mathrm{ke}(\mb{r})=\varepsilon_\mathrm{ke}^c(\mb{r})+\varepsilon_\mathrm{ke}^i(\mb{r})=(1/2)|\mb{F}^c(\mb{r})|^2+(1/2)|\mb{F}^i(\mb{r})|^2\,.
\ee
We stress that $\mb{F}^i(\neq\sqrt{\rho}\mb{v}^i)$ and $\mb{F}^c(\neq\sqrt{\rho}\mb{v}^c)$ essentially contain similar information as the full wavefunction $\Psi$ since the Helmholtz decomposition for $\mb{F}$ also involves the spatial variation of the density. Furthermore, as already shown in~\citep{Liu2023}, the characteristic scale of the spectrum of the irrotational $\varepsilon_\mathrm{ke}^i(\mb{r})$ is directly related to the sizes of the granules, as further discussed in section \ref{sec:energy}.

\section{ Gravitationally bound solitons in Self-Interacting FDM} \label{sec:analytical}

A distinct characteristic of FDM halos is the presence of a solitonic core, which features Brownian-like motions~\citep{Schive2020PhRvL.124t1301S,Li2020,DuttaChowdhury2021,Zagorac2022} and oscillations without deformation of its shape \citep{Guzman:2006yc,Chavanis2011, Marsh2019, PhysRevD.103.123551, Chiang2021}.
In the limit of $g=0$, an empirical profile was proposed by~\citep{Schive2014,Schive2014a}, and on the other hand in the limit of $E_g\gg E_{\nabla^2}$, the Thomas-Fermi~(TF) approach gives an analytical solution for the gravitationlly bound solution of the GPPE, $\rho_\mathrm{TF}(r)=\rho_cr_\mathrm{TF}\sin(\pi r/r_\mathrm{TF})/\pi r$ with $r_\mathrm{TF}=\pi\sqrt{g/4\pi Gm}$. Both serve as good approximations for the numerical ground solution up to $\mathcal{O}(r_c)$~\citep{Indjin2024} with a characteristic core radius $r_c$ and are sufficiently good to describe the core embedded in a halo.
For $g=0$, the core mass $M_c$ and core radius $r_c$ (defined at half-density) are related by $r_c \propto \hbar^2/Gm^2M_c$~\citep{Membrado:1989bqo}. In contrast, $r_c$ in the TF regime is independent of $M_c$: $r_c\propto\sqrt{g/Gm}$ \citep{1939isss.book.....C, ODell:1999ewx, Goodman:2000tg,Peebles:2000yy,Bohmer2007, Chavanis2011,Rindler-Daller:2011afd}. Some details on the transition between these two limits are given below.

Here, to quantify the core properties we consider a general density profile of the form
\be
\rho = \rho_c \, \varphi\left(r/r_c,\, g/g_\ast\right)
\ee
where $\rho_c$ is the central density, $r_c$ a characteristic length scale for the location of half central density and  $g_\ast$ a characteristic interaction strength discussed below.  
Building on \citep{Chavanis2011},~\citep{Indjin2024} considered the impact of a non-zero, repulsive self-interaction on the shape of a virialized, FDM soliton. The latter was characterized via 5 $g$-dependent {\em shape parameters}, corresponding to dimensionless integrals associated to the soliton's mass, different components of the soliton's energy, and its moment of inertia. These shape parameters, denoted by $\eta_g$, $\sigma_g$, $\nu_g$, and $\zeta_g$ are defined through their relevant integrals as
\begin{align}
    M_c &= \displaystyle4\pi \int_0^\infty \rho r^2 \; dr = \eta_g \, \left( 4 \pi \rho_c r_c^3 \right) \, ,
\end{align}
\begin{align}\label{eq:Quant-press}
    \Theta_Q &=\displaystyle \frac{2\pi \hbar^2}{2m} \int_0^\infty \left|\frac{\partial}{\partial r} \sqrt{\rho}\right|^2 \; dr = \sigma_g \left( \frac{\hbar^2 M_c}{m^2 r_c^2} \right) \, ,
\end{align}
\begin{align}\label{eq:Grav-Energy}
    W &\displaystyle= \frac{8\pi G}{2} \int_0^\infty r M_c(r) \rho \; dr = -\nu_g \, \left(\frac{G M_c^2}{r_c} \right) \, ,
\end{align}
\begin{align}\label{eq:Self-Energy} 
    U &= \displaystyle\frac{2\pi g}{m} \int_0^\infty \rho^2 r^2 \; dr  
    = \zeta_g \, \left( \frac{M_c^2 g}{2 m r_c^3} \right) \, .
\end{align}
A further shape parameter $\alpha_g$ associated with the moment of inertia
\begin{align}
    I &= 4\pi \int \rho r^4 dr = \alpha_g\, \left(M r_c^2\right).
\end{align}
appears in the study of small radial oscillations of the soliton. Note that all shape parameters are functions of g, with their value depending on the particular shape $\varphi$ of the profile.  

To characterise the relative role of interactions, we introduce a characteristic interaction strength $g_*$, identified as the value of $g$ at which
the self-coupling energy  \eqref{eq:Self-Energy} and the quantum pressure energy \eqref{eq:Quant-press} give the same contributions to the energy budget 
\begin{align}\label{eq:Quant-Eq-Self}
    \Theta_Q(g_\star) = U(g_*)\,. 
\end{align}
Since the shape parameters $\nu_g$ and $\zeta_g$, as well as $r_c$ are functions of $g$, Eq.~\eqref{eq:Quant-Eq-Self} leads to an implicit relation that can be solved for $g_\star$ only numerically, and only after the functions $\nu_g=\nu(g)$, $\zeta_g=\zeta(g)$ etc.~have been determined. Although this may be feasible, we can avoid this complexity and still get a very good analytical estimate for the value of $g_\star$ at which the dominant non-gravitational effect switches from the pure wave contribution to the non-linear self-coupling one, by evaluating the shape parameters from the (empirical) non-interacting soliton profile~\citep{Schive2014,Schive2014a}
\begin{equation}\label{eq:empirical}
\rho_\mathrm{soliton}(r)=\rho_c[1+\lambda (r/r_c)^2]^{-8}
\end{equation}
of total mass $M_c$ (where $\lambda = 2^{1/8} - 1 \approx 0.091$). After using the virial theorem: $2\Theta_Q+W+3U=0$, or $W+5U=0$ after using \eqref{eq:Quant-Eq-Self}, we obtain a characteristic interaction strength $g_*$ in the form 
\begin{equation}
   g_* \equiv \left( \frac{\sigma_0^2}{\zeta_0 \nu_0} \right) \frac{10 \hbar^4}{ G M_c^2 m^3} \;, \label{eq:g_*}
\end{equation}
where $\sigma_0$, $\zeta_0$ and $\nu_0$ are the shape parameters computed from the $g=0$ empirical profile~\eqref{eq:empirical}. 

Because $g_\star$ also depends on the soliton's mass $M_c$, it becomes clear that the influence of self-interactions is not a direct function of $g$, but rather of the ratio $g/g_*$ - see \citep{Chavanis2011, Rindler-Daller:2011afd} for an early introduction and use of a related quantity. We thus define the relative interaction strength parameter\footnote{In \citep{Indjin2024} we used the notation $\Gamma$.}
\be
\Gamma_g=\frac{g}{g_\ast} \;.
\ee
Importantly, $\Gamma_g$ takes different values for different solitons, even for the same value of $g$. These different solitons are stabilized against gravitational collapse by the quantum pressure for $\Gamma_g \ll 1$ and the self-interactions for $\Gamma_g \gg 1$ , with $\Gamma_g \sim O(1)$ indicating an intermediate, transition region of ``moderate" self-coupling.

\begin{figure}
\centering
\includegraphics[width=\columnwidth]{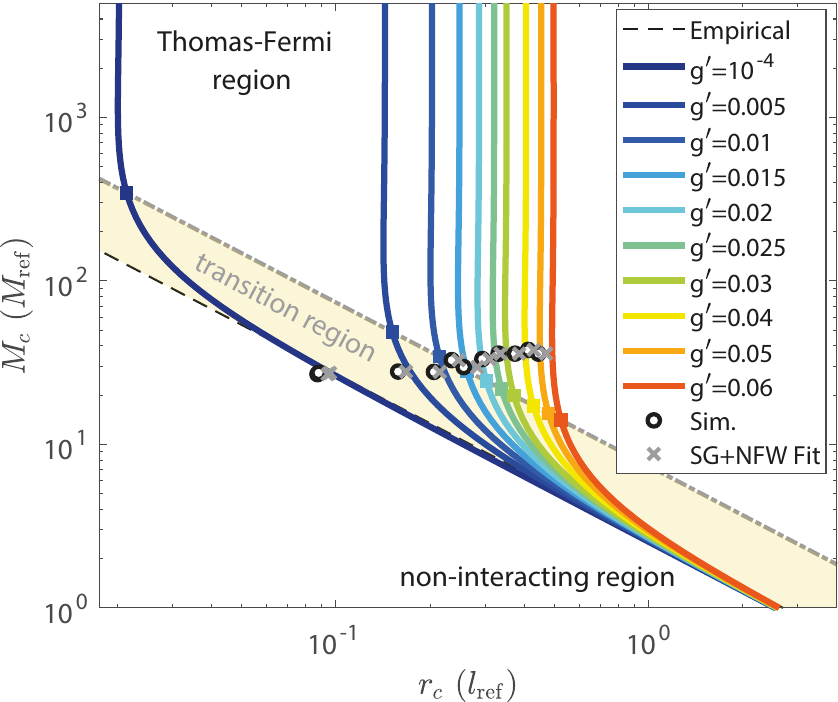}
    \caption{ 
    Relation between soliton core mass $M_c$ and radius $r_c$ clearly showcasing (i) the deviation of SFDM (curved solid lines) from FDM (dashed diagonal black line) and (ii) the regime probed in this work (circles/crosses) extending from the non-interacting ($g=0$) to the moderately-interacting $\Gamma_g \sim {\rm few}$.
    Presented analysis is based on \eqref{eq:radius_g},
with the shape parameters computed from the SG profile, Eq.~(\ref{eq:superGauss}), for the dimensionless interaction strengths $g^\prime$ considered in this work -- see section~\ref{sec:Numerics} for definitions. Along each solid line, $\Gamma_g$ increases as one moves from the bottom right (non-interacting solitons with $\Gamma_g\ll 1$) to the top (strongly-interacting Thomas-Fermi solitons with $\Gamma_g\gg 1$), with the indicative crossover value $\Gamma_g=1$ labeled by the squares. The latter are connected by the grey dash-dotted line while the black dashed line marks the non-interacting $g=0$ case, computed here from the empirical profile \eqref{eq:empirical}. The halo cores produced in our numerical mergers are also indicated with their masses $M_c$ obtained by fitting the SG profile and the core radii $r_c$ computed both from the SG+NFW fit \eqref{eq:core-halo} (grey crosses) and by directly interpolating the position of half peak density (black circles), showing the general consistency of their determination from both methods. 
    }\label{fig:2}
\end{figure}

Setting $dE/dr_c=0$ leads to a relation equivalent to the demand that the Virial theorem is obeyed~\citep{Chavanis2011,Rindler-Daller:2011afd}, namely 
\be
-2E_\mathrm{qp}-3E_g=E_\Phi\;.
\ee 
This relation can then be re-arranged to provide an expression for the characteristic solitonic core radius of the self-interacting system in terms of the boson mass $m$ and the soliton's mass $M_c$. We find
\begin{equation}
    r_c(\Gamma_g) = \frac{\sigma(\Gamma_g)}{\nu(\Gamma_g)}\frac{\hbar^2}{G M_c m^2} \left(1+\sqrt{1+15\,\mathcal{C}(\Gamma_g)}\right)\label{eq:radius_g}
\end{equation}
where we have introduced a generalised dimensionless interaction strength parameter $\mathcal{C}(\Gamma_g)$ which also accounts for interaction-induced shape effects via
\begin{equation}
\mathcal{C}(\Gamma_g) \equiv \left[ \frac{\zeta(\Gamma_g)}{\zeta_0} \,\, \frac{\nu(\Gamma_g)}{\nu_0} \,\,\left(\frac{\sigma^2(\Gamma_g)}{\sigma^2_0}\right)^{-1} \right]\,\,\Gamma_g \;.
\label{eq:Cofg}
\end{equation}
The $M_c-r_c$ relation is plotted in Fig.~\ref{fig:2}, clearly showcasing the deviation from the $g=0$ FDM prediction of
$M_c \propto 1/m^2 r_c$ 
(diagonal black dashed line), leading to the $M_c$-independent value
$r_c\propto\sqrt{g/Gm}$ (thick solid lines), which depends on the ratio $(g/m)$ of the two fundamental parameters of SFDM. 
The corresponding dependence of the central density on interactions is then given by 
\begin{align}
   \rho_c(\Gamma_g)  = \frac{1}{4\pi\eta(\Gamma_g)}\, \frac{M_c}{r_c^3(\Gamma_g)} \,.
   \label{eq:rho_0_g}
\end{align}
Finally, the frequency of oscillations of the interacting system \citep{Chavanis2011, Indjin2024} can be expressed in terms of $\Gamma_g$ via
\begin{eqnarray}
    f(\Gamma_g) &= & f(0)\left( \frac{ 2 }{ 1+\sqrt{1+15\,\mathcal{C}(\Gamma_g)} }\right)^2  \nonumber \\
    &\times & \sqrt{1+\frac{15\,\mathcal{C}(\Gamma_g)}{1+\sqrt{1+15\,\mathcal{C}(\Gamma_g)}}}
    \label{eq:gamma_freq}
\end{eqnarray}
where
\begin{equation}
    f(0) = \frac{\nu_0^2}{4\pi\sqrt{2 \alpha_0 \sigma_0}} \frac{G^2 M_c^2 m^3}{\hbar^3}. \label{eq:freq_mM}
\end{equation}
is the corresponding frequency of the non-interacting case.

In this work, we use an approximate analytical form for the soliton density profile in the form of a super-Gaussian (SG) 
\begin{equation}\label{eq:superGauss}
    \rho_{SG}(r) = \rho_c(\Gamma_g) \exp\left[-\ln 2 \left( \frac{r}{r_c(\Gamma_g)}\right)^{\vartheta(\Gamma_g)} \right] \;,
\end{equation}
which we have previously found to be an excellent fit for the cores of self-interacting FDM halos \citep{Indjin:2025lqs}. If $\Gamma_g$ is known for a given soliton, the exponent $\vartheta$ can be obtained from a semi-analytical expression 
\begin{equation}
    \vartheta(\Gamma_g) = \frac{\vartheta_0+\vartheta_{TF}}{2}-\frac{\vartheta_0 - \vartheta_{TF}}{2} \tanh\left( \frac{\log_{10}\left(\Gamma_g\right)-0.6}{1.5}  \right) \;, \label{eq:shape_fit}
\end{equation}
with $\vartheta_0 = 1.62 $ and $\vartheta_{TF} = 2.3 $ the exponent values at the non-interacting and TF limit respectively. The shape parameters are then computed using the profile, Eq.~\eqref{eq:superGauss}, (see Appendix~\ref{app:shape-params}).

In what follows we will use the above results to study the solitonic cores of halos formed in merger simulations including self interactions. The solitons in this work are characterized by $0\leq\Gamma_g\lesssim 6.3$, such that the examined cores range from weakly to moderately interacting ones. Such moderately-interacting values appear to be astrophysically relevant, as they are consistent with the inferred values for the most weakly-interacting  cores of dark-matter-dominated galaxies from the SPARC dataset (e.g. KK98-251) \citep{Indjin:2025lqs}\footnote{Note that our parallel study \citep{Indjin:2025lqs} showed the rotation curves of the most dark-matter-dominated SPARC database galaxies \citep{SPARC} to be well described by a {\em single} $m$ and $g$ combination, with $\Gamma_g \in [4.8, 1630]$.}.
As we will see, for this parameter range the effects of self-interactions will be visible in the properties of the cores but will not have correspondingly significant effects on the surrounding halos.

\section{Numerics}
\label{sec:Numerics}

In this section we describe the numerical procedure of our merger simulations.

\subsection{Reference Scales \& Dimensionless GPPE}
We numerically solve the GPPE in dimensionless format by scaling time, length and energy with the reference scales:
\be
\tau_\mathrm{ref}=(G\rho_\mathrm{ref})^{-1/2}\approx14.91\;\mathrm{ Gyr}\left(\frac{10^3M_\odot\mathrm{kpc}^{-3}}{\rho_\mathrm{ref}}\right)^{1/2},
\ee
\ba{rl}
l_\mathrm{ref}=&\displaystyle\left(\frac{\hbar\tau_\mathrm{ref}}{m}\right)^{1/2}
\\\\
\approx&\displaystyle10.81\;\mathrm{kpc}\left(\frac{2.5\times10^{-22}\;\mathrm{eV}}{mc^2}\right)^{1/2}\left(\frac{10^3M_\odot\mathrm{kpc}^{-3}}{\rho_\mathrm{ref}}\right)^{1/4},
\ea
and 
\be
E_\mathrm{ref}=\hbar\tau_\mathrm{ref}^{-1}\approx1.40\times10^{-33}\;\mathrm{eV}\left(\frac{\rho_\mathrm{ref}}{10^3M_\odot\mathrm{kpc}^{-3}}\right)^{1/2}
\ee
respectively. The dimensionless GPPE system is then written as
\be
i\frac{\partial}{\partial t^\prime}\Psi^\prime=\left[-\frac{\nabla^{\prime2}}{2}+g^\prime\left|\Psi^\prime\right|^2+\Phi^\prime\right]\Psi^\prime \;,
\ee
\be
\nabla^{\prime2}\Phi^\prime
=4\pi \varrho^\prime\left[\left|\Psi^\prime\right|^2-1\right] \;.
\ee
Here the wavefunction has been scaled by the averaged density of the system $\rho_\mathrm{sys}$, i.e.~$\Psi^\prime=\Psi/\sqrt{\rho_\mathrm{sys}}$; as a result the wavefunction is normalized to the simulation's box size,
and $\varrho^\prime=\rho_\mathrm{sys}/\rho_\mathrm{ref}$ is the ratio of the system to the reference density. 
We have introduced a dimensionless self-interaction strength $g'$, and correspondingly a dimensionless characteristic interaction strength $g'_\ast$ as  
\be
g^\prime= \left( \frac{\rho_\mathrm{sys}}{E_\mathrm{ref}} \right) \, g \, , 
\hspace{1.0cm}
g'_\ast= \left( \frac{\rho_\mathrm{sys}}{E_\mathrm{ref}} \right) \, g_\ast
\;, 
\ee
with
$\Gamma_g = g/g_\ast = g'/g'_\ast$.
The total energy can then be written as 
\be
E=\displaystyle\mathcal{E}_\mathrm{ref}\varrho^\prime\int d\mathbf{r}^\prime\Psi^{\prime\ast}\left[-\frac{\nabla^{\prime2}}{2}+\frac{g^\prime}{2}\left|\Psi^\prime\right|^2+\frac{\Phi^\prime}{2}\right]\Psi^\prime \;,
\ee
where 
\ba{rl}
\mathcal{E}_\mathrm{ref}=&\displaystyle\frac{E_\mathrm{ref}M_\mathrm{ref}}{m}=\frac{\hbar^{5/2}G^{-1/4}\rho_\mathrm{ref}^{3/4}}{m^{5/2}}
\\\\
\approx&\displaystyle1.26\times10^{42}\,\mathrm{\frac{kg\cdot m^2}{s^2}}
\left(\frac{\rho_\mathrm{ref}}{100 M_\odot\mathrm{kpc}^{-3}}\right)^{3/4}
\\
&\displaystyle\times
\left(\frac{mc^2}{2.5\times10^{-22}\mathrm{eV}}\right)^{-5/2} 
\ea
is a scale representative of the total system energy, as opposed to $E_{\mathrm{ref}}$ which represents the energy of a single particle. The dimensionless GPPE is solved using commonly-adopted numerical algorithms -- see, e.g.,~\citep{Schive2014, Schive2014a, Mocz2017, DuttaChowdhury2021, Liu2023, Alvarez-Rios2023} and references therein.

In this work, we focus on the role of self-interactions on the emerging density profiles, and their variable degrees of coherence across the entire cored-halo domain. As such, our analysis builds upon the $g=0$ case analyzed in \citep{Liu2023}. We thus consider here a total halo mass $M=100M_\mathrm{ref}$, where 
$M_\mathrm{ref}\equiv\rho_\mathrm{ref}12^3l_\mathrm{ref}^3 \approx1.26\times10^{6}M_\odot$ for the chosen reference values $mc^2=2.5\times10^{-22}$ eV and $\rho_\mathrm{ref}=10^3M_\odot\mathrm{kpc}^{-3}$.
The simulations in this work are performed with an improved\footnote{Note that this gives a spatial grid discretization of $\Delta x^\prime\approx0.03$ in this work, whereas the analysis in~\citep{Liu2023} was based on $\Delta x^\prime\approx0.069$ within a computational box size $(10l_\mathrm{ref})^3$ spanned by $288^3$ grid points, and $\varrho^\prime=0.1$.} spatial resolution in a slightly larger cubic computational box spanning $(12l_\mathrm{ref})^3$ with $384^3$ grid points, with $\varrho^\prime=100/12^3\approx0.058$.

\subsection{Merger Simulation}
\begin{figure*}
\includegraphics[width=2\columnwidth]{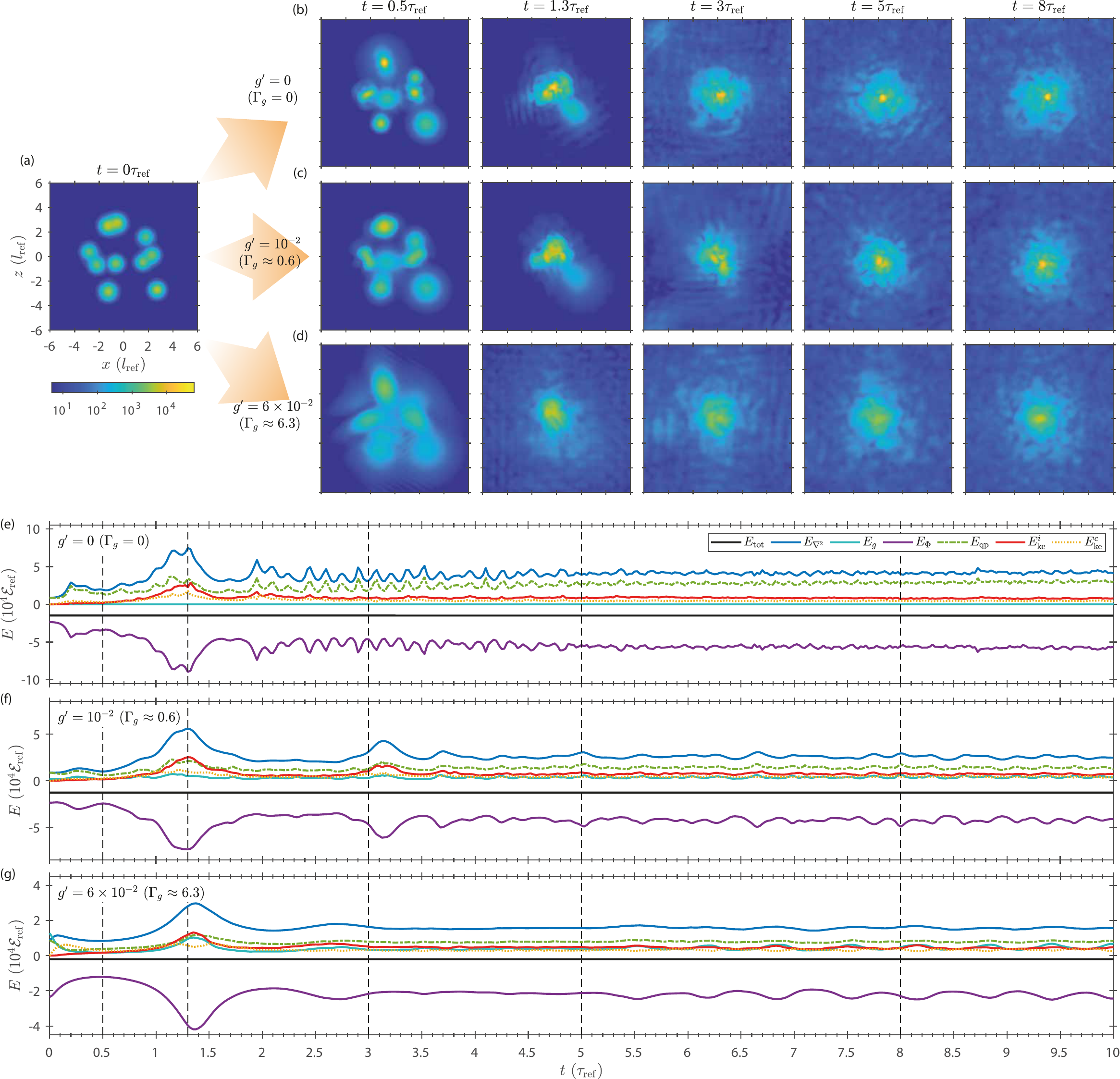}
    \caption{
    Initial conditions, merging volume rendering visualisations and energy dynamics for different self-coupling strengths.
    (a) The initial mass distribution condition ($t=0$), which is identical to the primary example that used in our previous $g=0$ work~\citep{Liu2023}. (b)-(d) Evolution snapshots at  $t/\tau_\mathrm{ref}=0.5$, $1.3$, $3$, $5$, $8$ are shown for $g^\prime$ (and corresponding $\Gamma_g$) values of 0 (0) (top), $0.03$ (0.6) (middle) and $0.06$ (6.3) (bottom). 
    (e)-(g) Corresponding energy evolutions (note the different vertical axes).
    The black vertical dashed lines mark the times of the density snapshots shown in (b)-(d).
    }\label{fig:3}
\end{figure*}

Here we focus on the investigation of a single dynamically equilibrated halo containing the core-halo structure of a dense and flattened core surrounded by a  Navarro-Frenk-White~(NFW) density profile. Similar to~\citep{Schwabe2016, Mocz2017, Chan2021, Liu2023, Alvarez-Rios2023, RindlerDaller2025}, our halo is generated by the gravitational coalescence  of several density lumps.
For consistency with our previous work~\citep{Liu2023}, the initial condition of our simulations is based on a configuration of 10 randomly distributed Gaussians with slightly different masses and widths set by their virial radii at $g=0$   \citep{Chavanis2011}, as shown in Fig.~\ref{fig:3}(a).
Our simulation is subject to a periodic boundary condition imposed by the implementation of the Fourier transformation, thus
allowing  particles to fly out, and back in, at the edges of the box~\citep{Alvarez-Rios2023}. This should be a relatively realistic boundary condition for big enough boxes with sufficiently low density at the edges.

Characteristic dynamical `virialization' snapshots are shown in Fig.~\ref{fig:3}~(b)-(d) for different (increasing) self-interaction strengths: Specifically, Fig.~\ref{fig:3}~(b) shows the non-interacting limit\footnote{This case is closely connected to the primary analysis of~\citep{Liu2023}.}, against which all our interacting results are to be contrasted, with Fig.~\ref{fig:3}~(c)-(d) showing, on the same timescales, the evolution in the presence of repulsive interactions (which clearly enhance the timescales of the initial density mixing). Panels (b)-(d) respectively correspond to dimensionless self-interaction values $g^\prime=0$, $0.01$ and $0.06$, which here result in solitons with $\Gamma_g=g/g_\ast \approx 0$, $0.6$, and $6.3$, implying that our probed interactions lie in the weak to moderate interaction range for the studied configuration.

The corresponding energy component evolutions, revealing the dynamical equilibration, are shown in Fig.~\ref{fig:3}~(e) to (g), with the vertical dashed lines indicating the times at which the snapshots shown in panels (b)-(d) were taken.
Consistent with our earlier non-interacting analysis~\citep{Liu2023}, we find that even in the presence of interactions (at least for the $\Gamma_g\in[0,6.3]$ values probed), the dynamical merger evolution is composed of three main stages, namely: (i) an initial early-stage drastic energy exchange in the kinetic and potential energy; (ii) a subsequent stage of non-negligible exchanges between the different energy components; and, finally, (iii) the emergence of practically (dynamically) virialized states, in which all energy components are effectively (on average) constant, barring some small-amplitude oscillations and an (anticipated \citep{Chen2021}) very slow change associated with an increase in the central density. It is worth noting here that, as $g^\prime$ increases, the oscillation period in energies at stage (ii) has a trend of increasing. However, our results are based on time-averaging and are not sensitive to such oscillations.
Confirmation of such virialization within our numerical timescales even in the presence of interactions, allows us to consider the emerging cored-halo profiles, and thus analyze their coherent and incoherent features.

\section{SFDM Halo Density Profiles and their Coherence} \label{sec:profiles}

Extending our previous work~\citep{Liu2023}, we now study the distinction between the highly-coherent solitonic core in the inner region, and the granule-filled vortex-turbulent state (`quasi-condensate') occupying the surrounding region, focusing on how self-interactions modify these substructures compared to the $g=0$ case.
The numerical radial density profiles 
$\rho_\mathrm{avg}(r)=\sum_{r\leq|\mb{r}|<r+\Delta  r}\rho_\mathrm{avg}(\mathbf{r})/\mathcal{V}_r$ with the shell volume $\mathcal{V}_r$ are obtained by angular averaging and also appropriate numerical temporal averaging for the near-steady states. Although only marginal energy exchanges take place between energy components after $t\approx5\tau_\mathrm{ref}$, see Fig.~\ref{fig:3}, it takes longer for residual transient oscillations, excited from the emerging dynamics, to become sufficiently subdued. In addition, the well-documented slow growth of the core mass on a longer time scale~\citep{Chen2021} precludes us from conducting long time averaging. Therefore, we consider a fixed averaging time span $t_\mathrm{avg}=3.5\tau_\mathrm{ref}$ for all our simulations, and average over 140 snapshots over such period. The averaging commences after we have ensured that near-steady states have been reached (see Appendix~\ref{sec:appendix-b}): such a time is identified by quantifying the  mass enclosed within $r_c$ (obtained, through interpolation, from the location of the half peak density) and ensuring it shows minimal fluctuations -- see Fig.~\ref{fig:appendix}.

The time-averaged radial density profiles are illustrated
for different (increasing)  interaction strengths $g^\prime=0$, $0.01$, $0.03$ and $0.06$ 
by the solid black lines in panels (a)(i)-(d)(i) of Fig.~\ref{fig:4} respectively. Extending beyond the usual $g=0$ FDM profiles comprising of an empirical solitonic core fit in the inner region and an NFW profile in the outer region, our $g >0$ spherically averaged numerical profiles can be well-approximated by the bimodal fitting function \citep{Indjin:2025lqs}
\be\label{eq:core-halo}
\rho_\mathrm{c-h}(r)=\left\{\begin{array}{ll}
     \rho_\mathrm{SG}(r) &  r\leq r_t\\
     \rho_\mathrm{NFW}(r) &  r > r_t
\end{array}\right. \;.
\ee
Here the SG profile is given in \eqref{eq:superGauss} and the well known  NFW profile is given by
\be
\rho_\mathrm{NFW}(r)=\rho_h\left(\frac{r}{r_h}\right)^{-1}\left[1+\left(\frac{r}{r_h}\right)\right]^{-2}
\label{eq:halo}
\ee
featuring the halo length scale $r_h$ and showing the expected  $r^{-3}$ trend at large $r$. The continuity of $\rho$ at $r_t$ limits the number of free parameters, determining the value of $\rho_h$.

To determine the best fitting bimodal  function for our halos, we perform a grid search in the order $M_c \rightarrow r_t \rightarrow r_h$, varying these parameters in the range $ M_c \in [10 M_{\rm ref}, 100 M_{\rm ref}]$ (appropriate for our configurations), $r_t \in [r_c, 10r_c]$ and $r_h \in [0.05 r_t, 5 r_t]$. Specifically, with $m$ and $g$ given, a value of $M_c$ fixes $\Gamma_g$ and hence $\vartheta$ via \eqref{eq:shape_fit}. From this, one is able to compute the shape parameter values stemming from the Super-Gaussian -- see formulae in Appendix \ref{app:shape-params}. This in turn enables the calculation of the soliton parameters, $r_c$ and $\rho_c$ via Eqs. \eqref{eq:radius_g}, \eqref{eq:Cofg} and \eqref{eq:rho_0_g}. Each resulting bimodal $\rho_{c-h}(r)$ profile is compared to the numerical density profile and evaluated for goodness-of-fit via a $\chi^2$ selection criterion similar to those employed in \citep{Indjin:2025lqs}. The best fitting profile for each $g'$ yields very good agreement in the overall shape of the spherically averaged halo density of our GPPE numerical simulations, confirming our procedure.

\subsection{Core-halo Profiles}
\begin{figure*}
 \centering
	\includegraphics[width=1.8\columnwidth]{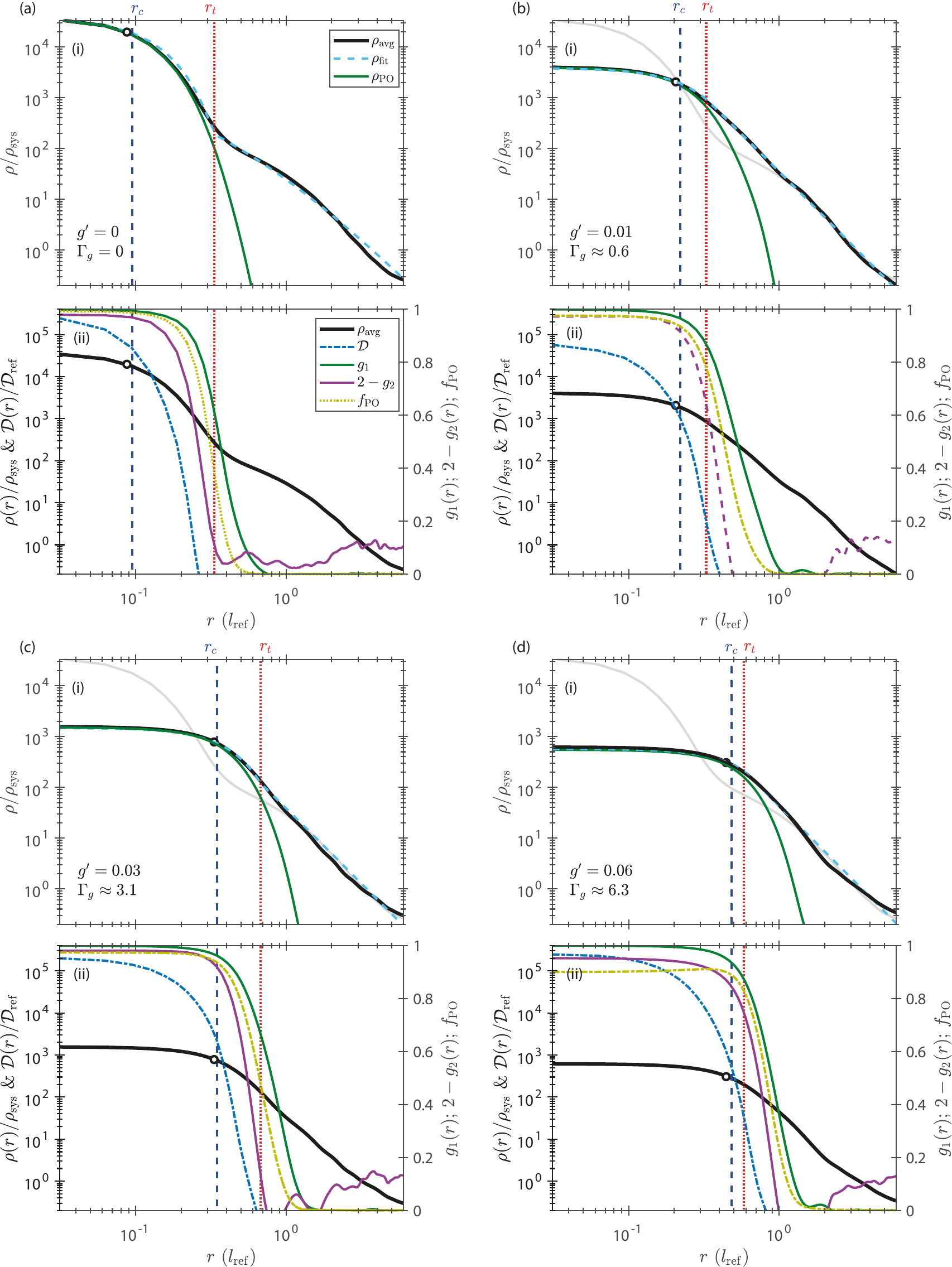}
    \caption{Core-halo profiles (i) and characterization of core-halo coherence (ii) for (a) non-self-interacting FDM halo, $g^\prime=0$ ($\Gamma_g=0$) and self-interacting ones with (b) $g^\prime=0.01$ ($\Gamma_g\approx0.6$), (c) $g^\prime=0.03$ ($\Gamma_g\approx3.1$) and (d) $g^\prime=0.06$ ($\Gamma_g\approx6.3$).
    Panel (i) and (ii) respectively illustrate different condensation and coherence measures, highlighting the crossover from coherent to incoherent field configurations as a function of radius $r$ (plotted in units of $l_{\rm ref})$: (i) [left axis] the time-averaged density profile $\rho_\mathrm{avg}$ (solid black), core-halo fit $\rho_\mathrm{fit}$ (dashed cyan) and $\rho_\mathrm{PO}$ (solid green) plotted in scaled units $\rho/\rho_\mathrm{sys}$ and $r/l_\mathrm{ref}$; (ii) shows the scaled phase space density $\mathcal{D}/\mathcal{D}_\mathrm{ref}$ (blue dash-dotted, left axis), and (right axis) the first-order correlation function $g_1$ (solid green), the density correlation function $g_2$ plotted as $(2-g_2)$ (solid purple), and the condensate fraction $\rho_\mathrm{PO}(r)/\rho_\mathrm{avg}(r)$ (yellow dash-dotted); the latter clearly reveals the deviation from a pure condensed fraction in the inner solitonic region due to interactions.
Note that, for clarity of the effect of repulsive interactions, plots (b)(i)-(d)(i) also show the $g=0$ non-interacting total density profiles by solid gray lines. The circles mark the interpolated location of the half peak density, while the blue dashed lines are obtained from the SG+NFW fit.
    \label{fig:4}}
\end{figure*}

The bimodal fits can be seen by the cyan dashed lines superimposed on the numerical densities in Figs.~\ref{fig:4}~(a)(i)-(d)(i). They uniquely identify the soliton and its spatial extent ($r_c$), the outer crossover region ($r_t$) and the NFW profile length scale ($r_h$). The values of $r_c$ and $r_t$ are marked by the vertical blue dashed and red dotted lines respectively. Clearly, they also allow for the ``extraction" of the soliton from the halo in which it is embedded.

The resulting SG solitonic core parameters and scaled interaction strengths $\Gamma_g$ for our numerically generated cored-halos are shown in Table~\ref{tab:example_table}, and are also positioned in the $M_c-r_c$ diagram of Fig.~\ref{fig:2}. As expected~\citep{Chavanis2011, Indjin2024} the higher-density central solitonic core exhibited in the central 
(small $r$) region in the absence of interactions, gradually transitions, with increasing interactions, into a more extended core region with a lower central density.

\begin{table}
	\centering
	\caption{
    Table of values characterising our simulated cored-halos: Dimensionless self interaction $g'$ and their approximate dimensionless ratio $\Gamma_g=g/g_\ast$ based on the displayed solitonic cores of masses $M_c$, peak densities $\rho_c$ and core radii $r_c$, as extracted from the SG profiles of Eq.~\eqref{eq:superGauss}.
    }
	\label{tab:example_table}
	\begin{tabular}{ccccccc}
        \hline
		$g^\prime$ & $\Gamma_g$  & $M_c$ ($M_\mathrm{ref}$) & $\rho_c$ ($\rho_\mathrm{ref}$) & $r_c$ ($l_\mathrm{ref}$)\\
		\hline
		0      & 0      & 26.9 & $2198.1$ & 0.094\\
		0.0001 & 0.006  & 27.4 & $2164.2$ & 0.096\\
        0.005  & 0.32  & 27.9 & $453.1$  & 0.171\\
		0.01   & 0.63  & 27.7 & $223.0$   & 0.220\\
        0.015   & 1.301  & 32.7 & $183.5$   & 0.253\\
        0.02   & 1.43  & 29.7 & $110.7$   & 0.290\\
        0.025   & 2.25  & 33.2 & $98.0$   & 0.318\\
        0.03   & 3.14  & 35.9 & $83.4$    & 0.345 \\
        0.04   & 4.19  & 35.9 & $57.7$     & 0.396\\
        0.05   & 5.81  & 37.8 & $45.2$     & 0.441\\
        0.06   & 6.29  & 35.9 & $32.9$     & 0.483\\
		\hline
	\end{tabular}
\end{table}

\subsection{Penrose-Onsager Mode and Spatial Coherence \label{sec:coherence-gppe}}

{The description of FDM dynamics via the classical GPPE is amply justified due to the enormous phase space density  
\be\label{cond_crit-1}
{\cal D}(r)=\frac{\rho(r)}{m} \lambda_\mathrm{dB}^3
=\frac{h^3}{m^4} \frac{\langle\rho(r)\rangle}{\langle|{v}(r)|\rangle^3} ={\cal D}_\mathrm{ref}\frac{\langle\rho^\prime(r)\rangle}{\langle|{v}^\prime(r)|\rangle^3} \;,
\ee
where
\ba{rl}
{\cal D}_\mathrm{ref}=&\displaystyle\frac{h^3\rho_\mathrm{ref}}{m^4v_\mathrm{ref}^3}
\approx\displaystyle1.25\times10^{90}
\\\\
&\times\left(\frac{10^{-22}\textrm{eV}/c^2}{m}\right)^4\left(\frac{\rho_\mathrm{ref}}{10^3M_\odot\mathrm{kpc}^{-3}}\right)\left(\frac{250\mathrm{km/s}}{v_\mathrm{ref}}\right)^3\,.
\ea
However, as discussed in \citep{Liu2023} (see Sec.~3.1 there) this does not necessarily identify the classical field that solves the GPPE with a "condensate". The term "Bose-Einstein Condensate" typically refers to a state which not only exhibits a high number of particles per de Broglie volume but also a suppression of phase and density fluctuations~
\citep{huang2000statistical, Pitaevskii2003,2008bcdg.book.....P}. At equilibrium, a homogeneous condensate is characterized by the existence of off-diagonal long-range order (ODLRO), i.e.~a finite value of the one-body density matrix 
\be\label{eq:PO}
\varrho(\mb{r},\mb{r}^\prime)\equiv\frac{1}{m}\langle\Psi^\ast(\mb{r})\Psi(\mb{r}^\prime)\rangle
\ee
as $|\mathbf{r}-\mathbf{r'}| \rightarrow \infty$,
which is consistent with the mode having both suppressed density and phase fluctuations~\citep{Pitaevskii2003, leggett2006quantum}, a configuration that might also be loosely termed as "coherent". In our case where gravity acts as a "self-focussing force" leading to the breaking of translational symmetry, the long-range order need not extend to infinity. This is analogous to inhomogeneous finite-sized laboratory systems, e.g.~harmonically-confined ultracold atomic gases, at finite temperatures, for which the central coherent condensate is surrounded by an incoherent thermal cloud, with the condensate only spanning the entire (finite-sized) atomic cloud in the $T \rightarrow 0$ limit -- an analogy clearly pointed out in~\citep{Liu2023,Proukakis:2023nmm}. For a theoretical discussion of the importance of simultaneous suppression of density and phase fluctuations evidencing the emergence of a condensate, see, e.g.,~\citep{PhysRevA.59.4595,PhysRevA.66.013615, Cockburn2011,PhysRevA.87.063611, proukakis2023encyclopedia}.

In such cases (as also applicable to our current SFDM context), the condensate can be identified with the field's Penrose-Onsager (PO) mode which corresponds to the largest eigenvalue of the one-body density matrix $\varrho(\mb{r},\mb{r}^\prime)$~\citep{1956PhRv..104..576P,leggett2006quantum}. Mathematically, the dominant eigenvalues and corresponding eigenfunctions are found from the solution of
\be
\int {\rm d}{\bf r'} \varrho ({\bf r}, {\bf r'}) \Psi_n(\mb{r'}) =  N_{n} \Psi_n(\mb{r})
\ee
where the index $n$ labels the $n$-th eigenfunction $\Psi_n$ of $\varrho$. If one eigenvalue is dominant the system is said to exhibit a well-defined Bose-Einstein condensate, with $\Psi_0=\Psi_{_\mathrm{PO}}$ the condensate mode, and $N_\mathrm{PO}=N_0$ the condensate particle number.}

In our simulations the PO mode is extracted numerically through time-averaging over a timescale sufficiently longer than other system dynamical timescales: Note that the use of time averaging here, instead of the formal quantum state/ensemble average $\langle\ldots \rangle$ in \eqref{eq:PO}, is established practice in the literature of inhomogeneous laboratory condensates
~\citep{2005PhRvA..72f3608B,2008AdPhy..57..363B}. The time averaging involved in extracting the PO mode is performed once the post-merger configuration has settled sufficiently and a central core has formed with a core mass which does not fluctuate significantly; more details on the choice of such timescale can be found in appendix \ref{sec:appendix-b}.       

Beyond the existence of a (dominant) PO mode, the coherence associated with condensation can also be assessed by the normalized correlation functions
\be\label{eq:g1}
g_1(\mb{r})=g_1(\mb{0},\mb{r})=\frac{\langle\Psi^\ast(\mb{0})\Psi(\mb{r})\rangle}{\sqrt{\langle|\Psi(\mb{0})|^2\rangle\langle|\Psi(\mb{r})|^2\rangle}} \;,
\ee
and
\be\label{eq:g2}
g_2(\mb{r})=g_2(\mb{r},\mb{r},\mb{r},\mb{r})=\frac{\langle|\Psi(\mb{r})|^4\rangle}{\langle|\Psi(\mb{r})|^2\rangle^2}=\frac{\langle[\rho(\mb{r})]^2\rangle}{\langle\rho(\mb{r})\rangle^2} \;,
\ee 
respectively characterizing phase and density fluctuations. A condensate would be characterized by 
\be
g_1(\mb{r}) \approx \, g_2(\mb{r}) \approx 1 \;,
\ee
while regions where
\be
g_1(\mb{r}) \rightarrow 0\;, g_2(\mb{r}) \rightarrow 2
\ee
involve a spatio-temporally evolving field that is not coherent enough to be considered a condensate. Note that the value $g_2\simeq 2$ is anticipated for approximately Gaussian fluctuations. Again, expectation values are interpreted as time averages over sufficiently long timescales - see appendix \ref{sec:appendix-b} and \citep{Liu2023} for more details.   

In the analysis that follows we use both the PO mode and $g_1(\mb{r})$, $g_2(\mb{r})$ to identify the condensate in our simulations. In \citep{Liu2023} we showed that for $g=0$ the central soliton is essentially indistinguishable from the PO mode, with the surrounding outer halo being a largely incoherent field configuration. The surrounding halo's often mentioned granules were found to be patches of partial and fleeting coherence. As we will see below, this picture of a BEC-soliton embedded in a halo of semi-coherent field configurations broadly applies also for $g\neq 0$ but with some subtle differences: with $g\neq 0$, the core itself is now also influenced by above-the-condensate modes leading to a (slightly) reduced PO mode at the centre, a phenomenon termed condensate depletion and discussed in the next subsection.  

The non-interacting FDM solitonic core, commonly parametrized by the empirical profile \eqref{eq:empirical}, is reproduced here from~\citep{Liu2023} in Fig.~\ref{fig:4}~(a)(i) where the green solid line depicting the numerically-extracted PO (condensate) mode traces the overall system density excellently beyond the characteristic solitonic length scale $r_c$, almost up to the outer radius, $r_t$. 
As evident in Fig.~\ref{fig:4}~(b)(i)-(d)(i), such correspondence between the PO mode and the averaged density appears very good for $r\lesssim r_c$ even in the presence of the (moderate) interactions studied here. As we will discuss below however, there is in fact a subtlety that prevents the complete identification of the PO mode with the core density when $g\neq 0$ which we can detect in our simulations.   

Panels (ii) in Fig.~\ref{fig:4}~(a)-(d) show different measures characterizing the wavy and coherent nature of dark matter, as first reported in~\citep{Liu2023}. Simultaneous suppression of both phase and density fluctuations emerges within a spatial region exhibiting a constant value (plateau) with $g_1(\mb{r}) \approx \, g_2(\mb{r}) \approx 1$,
a condition evidently well-satisfied for $r \lesssim r_c$ for all probed interaction strengths. In the region $r_c \lesssim r \lesssim r_t$ the system transitions from a coherent behaviour (representing the solitonic core) towards an incoherent (chaotic) behaviour, as evident by the fact that  $g_1(\mb{r}) \rightarrow 0$ and $g_2(\mb{r}) \rightarrow 2$ (the latter anticipated for purely Gaussian fluctuations).

Another way to visualize the transition with radius from coherent to incoherent field is by considering the condensate fraction ($\rho_{_\mathrm{PO}} = |\Psi_{_\mathrm{PO}}|^2$)
\be
f_{_\mathrm{PO}}(r)\equiv\rho_{_\mathrm{PO}}(r)/\rho(r) \;,
\ee
whose radial dependence is shown by the dash-dotted yellow line in Fig.~\ref{fig:4}~(a)(ii)-(d)(ii).
The existence of a plateau with $f_{_\mathrm{PO}}(r)\approx 1$ for $r \lesssim r_c$ confirms that the solitonic core largely corresponds to a nearly pure condensate, with the condensate fraction rapidly decreasing at larger $r$, in agreement with the other coherence measures. However, as $g$ increases, $f_{_\mathrm{PO}}(r)$ decreases below 1 by a small amount within $r_c$, including deep within the solitonic region (i.e. as $r \rightarrow 0$) -- such condensate depletion is discussed in Sec.~\ref{sec:depletion}.

Beyond the inner solitonic core region, all our characterisation measures indicate a qualitative change in the system behaviour. Specifically, the system becomes dominated by the density fluctuations and coherence is lost, as evident from $g_1(r) \rightarrow 0$, $g_2(r) \rightarrow 2$ and $f_{\rm PO} \rightarrow 0$. Such regime is dominated by a tangle of quantum vortices, as discussed in Sec.~\ref{sec:energy}.

For completeness, we also show the numerical dimensionless phase space density ${\cal D}(r)$  by the blue dash-dotted lines in Fig.~\ref{fig:4}(a)(ii)-(d)(ii).
For our typical low boson masses, ${\cal D}_\mathrm{ref}$ assumes extremely large values, thus justifying the wave description used here over the entire simulation grid, even in the absence of a pure condensate towards the grid edges. Note however the change in behaviour of ${\cal D}(r)/{\cal D}_\mathrm{ref}$ as the density transitions to the NFW profile, also demarcating a qualitative difference between the condensed core and incoherent enveloping halo (occurring around the condensation transition region  ${\cal D}(r)/{\cal D}_\mathrm{ref} \sim O(1)$).

\subsection{Condensate Depletion of SFDM Halos \label{sec:depletion}}

Interestingly, our simulations exhibit a clear dependence of the inner solitonic core's level of coherence on the interaction strength:
Closer inspection shows that the Penrose-Onsager modes [solid green lines in Fig.~\ref{fig:4}~(a)(i)-(d)(i)] are located slightly below the total density profiles (corresponding solid black lines), although the logarithmic axis scale does not make this easily noticeable. The effect becomes more evident in the extracted condensate fraction dependence on radius [dash-dotted yellow lines in Fig.~\ref{fig:4}~(a)(ii)-(d)(ii)].
Such lines reveal a condensate fraction $f_{PO} < 1$, with the deviation from the fully-coherent value of 1 seemingly increasing over the entire soliton width with increasing $g$.
As argued below, we believe the dominant mechanism is the effective softening of the gravitational potential of the final steady-stated cored halos, as a result of the increase in the interaction energy of the initial state of our merger simulations.

To gain further insight on this observation, Fig.~\ref{fig:5} plots the {\em mean} condensate depletion 
\be
f_{\mathrm{PO},\,\mathrm{int}} = 4\pi\int_0^{R} drr^2\frac{\rho_{_\mathrm{PO}}(r)}{\rho_\mathrm{avg}(r)} \left/ \left(\frac{4\pi R^3}{3}\right)\right.
\label{eq:fPO_int}
\ee
integrated to different radii $R$ within the soliton as a function of the corresponding dimensionless self-coupling strength $g'$ and $\Gamma_g$.
Open circles show the results integrated up to the solitonic core region, $R= r_c$, with corresponding values obtained for $R=r_c/2$ (upper dashed line), $R = 1.2 r_c$ (dash-dotted line) and $R = r_t$ (dotted line) also shown.
These values of $f_{\mathrm{PO,\,int}}$ provide an effective error range for estimating $f_{\rm PO}$ in a situation where translational symmetry is broken and the PO mode is embedded in a more extended fluctuating field.
{The main features of this figure are: (i) the enhancement of condensate depletion as one increases the region of integration from the centre of the soliton [$R\rightarrow r_t$, i.e.~vertical change in $f_{\rm PO}$ at fixed $g'$], which can be attributed to the fact that the field exhibits more dynamical phase and density fluctuations as one moves from the core to the surrounding halo; (ii) the overall enhancement of depletion with increasing interaction strength $g'$ [horizontal change in $f_{\rm PO}$] compared to the non-interacting case, which however (iii) also exhibits an apparent non-monotonic behaviour around $\Gamma_g \sim 3$ (within the vertical yellow band).


\begin{figure}
 \centering
	\includegraphics[width=1\columnwidth]{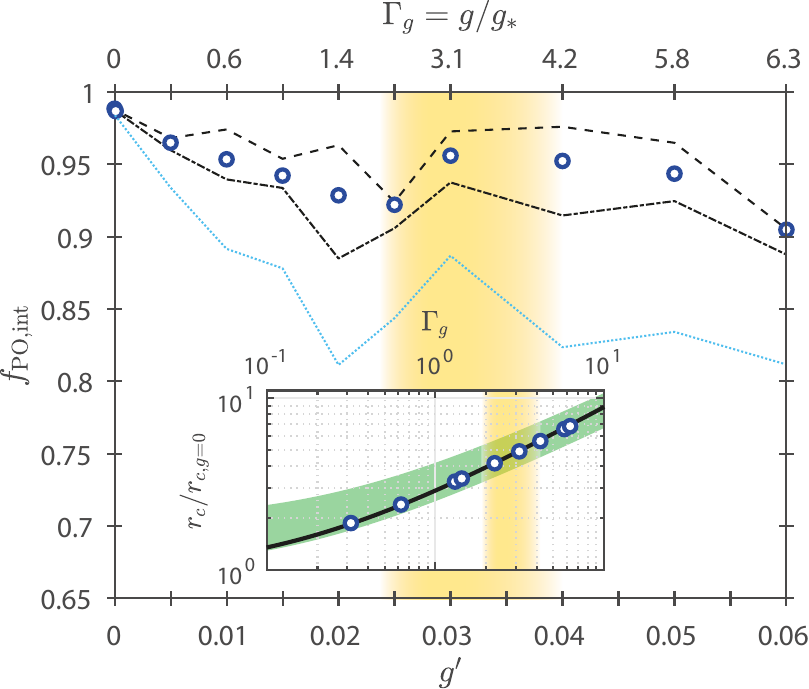}
    \caption{Depletion of the coherent core measured by the integrated Penrose-Onsager~(PO) condensate density fraction,  Eq.~\eqref{eq:fPO_int}
    , as a function of $g^\prime$ and $\Gamma_g=g/g_\ast$. The blue circles 
    denote the average fraction within the solitonic core $r_c$ (i.e.~$R=r_c$), while the black dashed and dashed-dotted lines consider $R=r_c/2$ and $R=1.2r_c<r_t$. The light blue dotted line is for $R=r_t$ which includes more fluctuations from the halo. The trend exhibited for increasing $R$ indicates the relevance of incoherent contributions to the depletion. 
    Inset: Theoretical prediction for $r_c/r_{c,g=0}$ with increasing $\Gamma_g$, reproduced from \citep{Indjin2024};
    the bottom and top sides of the green band show the prediction using $g=0$ and Thomas-Fermi limits of the soliton shape parameters respectively. The region $\Gamma_g\in(2,4)$, marked by the vertical yellow band, indicates a transition region where the shape of the soliton starts deviating appreciably from the $g=0$ formula \eqref{eq:empirical}. This region coincides with the apparent, locally non-monotonic behaviour of the otherwise downwards depletion trend with $\Gamma_g$.   
     }
    \label{fig:5}
\end{figure}

While a detailed explanation of the depletion and its observed trend with increasing $g'$ lies beyond the scope of the present paper, we nevertheless attempt a {\em qualitative} description of such emerging features {by drawing an {\em indirect} analogy between our steady-state SFDM cored halos and equilibrated inhomogeneous laboratory atomic BECs in external confining potentials (traps) at finite temperature.
In the latter systems, typically confined in a harmonic trap, the central highly-coherent condensate is surrounded by a cloud of incoherent (thermal) atoms, with the analogy to FDM clearly highlighted in Fig.~1 of~\citep{Proukakis:2023nmm} (based on simulations in~\citep{Liu2023}).

Firstly, we address the issue of condensate depletion in a weakly-interacting quantum gas. This
is associated with the promotion of bosons from the condensate ground state to low-lying excited states, and can occur in two different ways: either via virtual excitations arising as a result of quantum fluctuations, or due to real excitations when the system acquires sufficient energy to populate low-lying excited modes~\citep{2008bcdg.book.....P}. As a result, the latter effect increases significantly with enhanced thermal fluctuations, with both effects typically being at play in different proportions, with their origin often hard to disentangle (see, e.g.,~\cite{atomic-depletion}).

In the simplest limit of a $T=0$ contact-interaction homogeneous Bose gas in the weakly-interacting limit $n a_s^3 \ll 1$, the dominant contribution to the depletion arises purely from quantum fluctuations and is given by the analytical formula $1-f_\mathrm{PO} \approx O(1) \sqrt{n a_s^3}$, where $n$ is the number density and $a_s$ the s-wave scattering length\footnote{Note that the FDM self-coupling constant has been defined as $g=4\pi\hbar^2a_s/m^2$, with $|\Psi|^2$ corresponding to mass density in our GPPE (as opposed to number density usually employed in cold atomic systems), such that our definition involves one extra power of mass compared to the standard definitions.}~\citep{2008bcdg.book.....P}. For example, this has been experimentally verified to hold for $n a_s^3 \lesssim 0.03$ in the context of weakly-interacting homogeneous atomic gases \citep{Lopes_2017}, with such quantum depletion effects becoming more pronounced in long-range-interacting systems, such as dipolar condensates~\citep{Blakie_Depletion,Chomaz_2023} and liquid Helium~\citep{1956PhRv..104..576P,helium-depletion}. With respect to such quantum depletion, we note that SFDM is \emph{extremely} dilute in our parameter range\footnote{Incidentally, this amply validates the use of the GPPE to describe the system dynamics \citep{Rindler-Daller:2022fjt}.}:
\ba{rl}
na_s^3=&\displaystyle \frac{m^5 \left(E_\mathrm{ ref}\right)^3}{64\pi^3\hbar^6\left(\rho_\mathrm{sys}\right)^2}\left(g^{\prime3}\frac{\rho}{\rho_\mathrm{sys}}\right)=\displaystyle\frac{m^5G^{3/2}}{64\pi^3\hbar^3\rho_\mathrm{ref}^{1/2}\varrho^{\prime2}}\left(g^{\prime3}\frac{\rho}{\rho_\mathrm{sys}}\right)
\\\\
\approx&\displaystyle1.5\times10^{-189} \, \times \,\left(g^{\prime3}\frac{\rho}{\rho_\mathrm{sys}}\right)
\\\\
&\displaystyle\times\left(\frac{mc^2}{2\times10^{-22}\;\mathrm{eV}}\right)^{5}\left(\frac{\rho_\mathrm{ref}}{10^3\;M_\odot/\mathrm{kpc}^3}\right)^{-1/2}.
\ea
ruling out the relevance of any such (purely) zero temperature effect.

At a non-zero temperature (where presumably the indirect analogy to SFDM becomes more relevant due to the SFDM's significant kinetic energy), the propensity of particles to occupy excited modes in a  harmonically trapped quantum gas with contact interactions can be loosely characterised in terms of the available excitation energy of the system. A relevant quantity to characterize this effect is the ratio of such energy to the typical harmonic oscillator level spacing $\hbar \omega$.
The relevant excitation energies in such weakly-interacting gases are the interaction energy $g \rho$ (in the notation of this paper), and the thermal energy $kT$.  
When either, or both, ratios $(g\rho/\hbar \omega)$, or $(kT/\hbar \omega)$ become $\gtrsim O(1)$, for a given total energy and configuration, (excited) modes become energetically accessible, with their occupation leading to condensate depletion. The number of excited modes becoming relevant in fact depends on the values of such ratios, which can typically be $\gg 1$ in 3D systems, implying a significant number of non-condensate modes.

Compared to relatively weakly-interacting laboratory condensates, SFDM features an additional indirect long-range interaction, through the gravitational force. The net effect is an attractive long-range force leading to gravitationally virialized cored halos. While this could be vizualised as an extra long-range interaction (somewhat analogous, e.g., to long-range dipolar interactions~\citep{Chomaz_2023,proukakis2025dipolar}), it is perhaps easier for our present discussion to consider the role of gravity in the steady-state solitonic cores as providing an effective intrinsic (self-consistent) confinement -- in a manner qualitatively rather analogous to external harmonic traps. Taking the SFDM soliton density to be roughly constant at $\rho_c$, we can easily extract an effective harmonic trap frequency of $\omega_G \sim \sqrt{G\rho_c}$. Our simulations with increasing $g$ result in a significant decrease in $\rho_c$ of the steady-state solitonic cores, with the dominant effect of this resulting in a shallower gravitational potential: such an effect is analogous to opening up the harmonic trap. This results in reducing the energy gap between the energy levels (in an eigenmode expansion of the field) above the ground state, thus making [for a given system energy/configuration] more (excited) modes accessible\footnote{Such a picture appears consistent with the consideration of density contributions from the excited states calculated within the SPE~\citep{Lin2018,Yavetz2021,Zagorac2022,Zagorac2023,Chan2023,Zimmermann2025}, or GPPE~\citep{Salasnich2025}.}.

Although the SFDM system is not in thermal equilibrium, one could indirectly infer an {\em effective} (and non-zero) temperature, through the system's classical kinetic energy, with such quantity found to slightly reduce with increasing $g$ (although we note that, within our atomic gas analogy, the critical temperature would itself also decrease by a factor of $\omega$). However, as such classical kinetic energy is clearly  much less than the interaction energy (at least deep) within the solitonic core region, and given that the depletion for $g=0$ is practically negligible, we assume that the (dominant) mechanism for the observed depletion in our SFDM simulations likely arises from the interplay of the interaction energy and decrease in $\omega_G$ stemming from the gravitational potential becoming shallower.

In particular, we find that (with the exception of negligible values for $\Gamma_g \ll O(0.1)$), with $g$ increasing and $\rho_c$ decreasing, the values of $g \rho_c$ in fact remain, to first approximation, constant over all probed interaction strengths. At the same time the effective harmonic trap frequency of the gravitational potential within the solitonic core reduces (approximately linearly) with $g'$, by a factor of few tens. We therefore suggest that the emerging monotonic increase in the value of $(g \rho_0 / \hbar \omega_G)$ due to the gravitational `trap' softening is the likely dominant cause of the observed increase in the condensate depletion within the solitonic core\footnote{Naturally, the situation becomes very different outside the solitonic core, as the interaction energy in the outer halo becomes practically irrelevant, with the observed  excitations originating from the kinetic energy contributions, in a manner {\em analogous} to temperature-induced phase- and density-fluctuations in laboratory condensates.} with increasing interaction strength.

At this stage it is also pertinent to comment further on the increase of the magnitude of depletion with enlarged integration range $R$, as measured from the solitonic core centre. From inhomogeneous  atomic condensates, we know that the condensate fraction significantly decreases as one moves towards the outer regions of the trap: this is a direct consequence of the central region being dominated by the (coherent) condensate, with the outer regions dominated by the (incoherent) thermal cloud. Hence the larger the integration volume, the more incoherent particles are included, leading to a decrease in the condensate fraction over an increasing volume. This is exactly what the different lines in Fig.~\ref{fig:5} show for the SFDM content, namely that condensate fraction decreases with increasing integration volume at a fixed interaction strength -- with this effect evidently becoming more dominant once regions outside the coherent solitonic core ($R \lesssim r_c$) are included [see, e.g.~ the significant decrease in the bottom dotted blue line, extracted at $R=r_t > 1.2 r_c$].
Physically, this can be attributed to the fact that the field exhibits more dynamical phase and density fluctuations as one moves from the core to the surrounding halo. }

Finally, we comment briefly on the emerging shape of the condensate depletion with increasing interaction strength, which exhibits a common feature irrespective of the integration range $R$: Although the above {\em qualitative} argument alone would suggest a monotonic trend in condensate depletion as a function of interaction strength, our numerical findings reveal a more complicated behaviour, in the form of a local feature in the trend of Fig.~\ref{fig:5} around $\Gamma_g \approx 3$ ($g' \lessapprox 0.03$). Our previous analysis in \citep{Indjin2024} has quantified how the solitonic profile starts to noticeably transition from the empirical $g=0$ profile \eqref{eq:empirical} to a Thomas-Fermi profile ($\Gamma_g \gg 1$) at around that value of $\Gamma_g$. This transition happens gradually from the non-interacting regime (generally valid for $\Gamma_g \lesssim 10^{-1}$) to the Thomas-Fermi regime (approached from around $\Gamma_g \gtrsim 10^2$), and the value $\Gamma_g \approx 3$ ($g' \lessapprox 0.03$) is quoted as the approximate point where various soliton properties, such as the central density $\rho_c$ or radius $r_c$ cross a midpoint between the non-interacting and the Thomas-Fermi solitons.
This was shown clearly in Fig.~3(b)(i) of~\citep{Indjin2024}, with the most relevant feature related to such a transition in the shape of the density reproduced here in the inset to Fig.~\ref{fig:5}.} 
Interestingly, as the observed weakly non-monotonic behaviour of the depletion in Fig.~\ref{fig:5}  occurs in the region around  $\Gamma_g \approx 3$, we infer that it could possibly be attributed to the local change in gravitational potential associated with the shape change of the soliton. 

Although the above {\em indirect} analogy gives some {\em qualitative} insight into the aspects of the observed behaviour,
a more detailed study would be required to explain the shape and magnitude of the observed drop in $f_\mathrm{PO}$.
This would require at least
 a more in depth analysis of the numerical GPPE solutions, but
possibly further necessitate the use of the self-consistently coupled coherent-incoherent dynamics, using for example the formalism of ~\citep{Proukakis:2023nmm,Proukakis:2023txk,Proukakis:PhysRevD.111.023505} within the core region. Such an analysis is however beyond the scope of this paper.

\begin{figure}
 \centering
	\includegraphics[width=1\columnwidth]{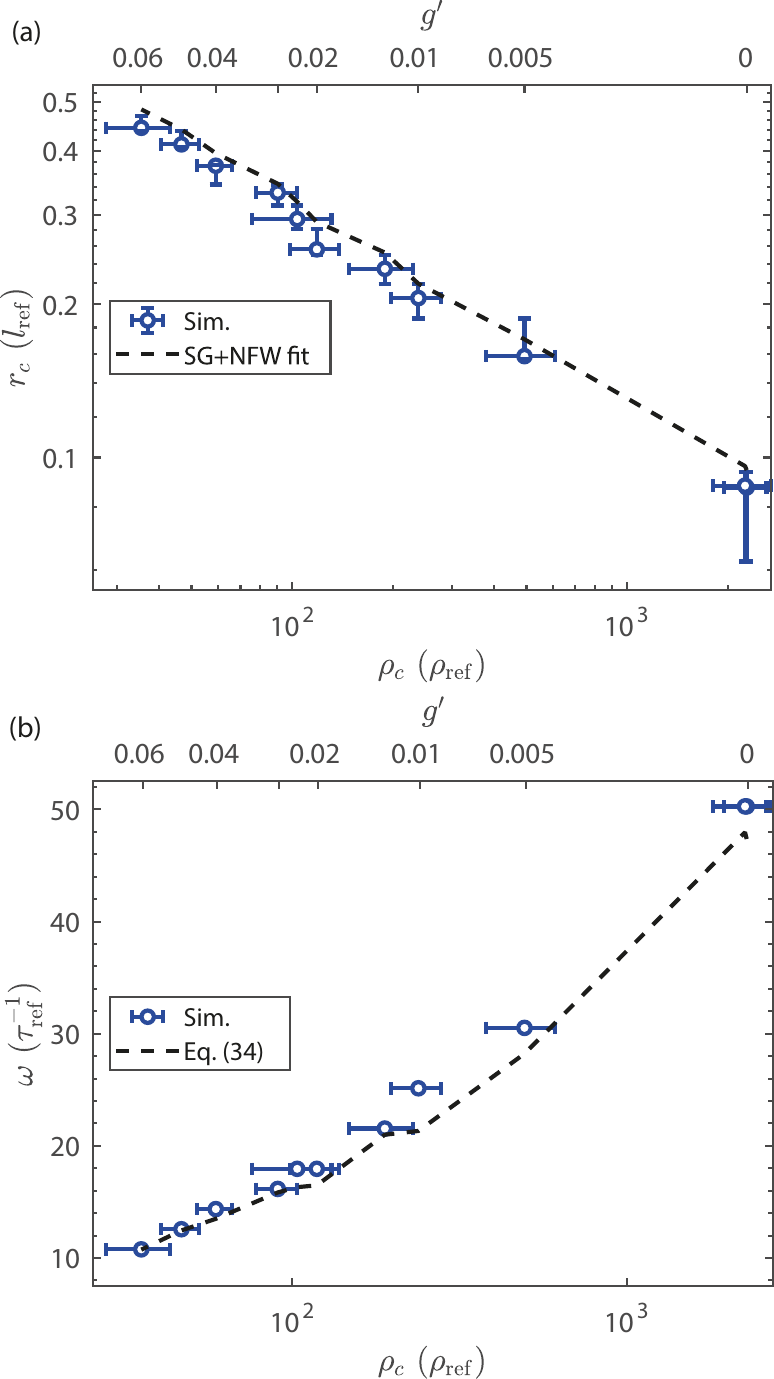}
    \caption{ (a) The relation between soliton $r_c$ and core density $\rho_c$. The dashed line represents Eq.~\eqref{eq:rho_0_g} with the core mass $M_c$ and the shape parameters computed using a best fit bimodal SG+NFW  profile, as explained in section \ref{sec:profiles}. The open circles denote the central density and soliton size ($r_c$ is the interpolated radius of half density), extracted directly from the numerical profiles. (b) The soliton oscillation frequency as a function of its central density. The dashed line represents Eq.~\eqref{eq:gamma_freq} ($\omega=2\pi f$) and the open circles are obtained from the temporal Fourier transform of $\rho_c(t)$ measured from our simulations during a time period where the core has settled to a quasi-equilibrated state (see Fig.~\ref{fig:appendix}). The vertical error bars in (a) express the confidence range of the location of the half-density point around the average in the oscillating core profile. The horizontal error bars in (a) and (b) express the standard deviation of the core density. The error in the peak location of the Fourier spectrum of the core density evolution is negligible. 
    }
    \label{fig:6}
\end{figure}

\subsection{Other Solitonic Core Properties: $\rho_c-r_c$ relation  and dominant oscillation frequencies}

The previous subsection showed that the halo's core is mostly made up of the field's PO (condensate) mode, but with a small addition of above-the-PO modes, evident in $f_{_\mathrm{PO}}<1$.  Distinct from the numerical PO mode analysis, the numerical results obtained by studying isolated solitons with $g \neq 0$ in \citep{Indjin2024} suggest that we can also use the SG analytical density profile \eqref{eq:superGauss} to separate out the soliton configuration from the total SFDM field of the halo, and compare theoretical predictions for isolated solitons with corresponding measurements for the core embedded in our simulated halos. 

In particular, we measure the central density ($\rho_c$)  and, independently, the inferred core size ($r_c$) can be determined via $\rho(r_c)=\rho_c/2$. These data points obtained from the numerical simulations are plotted in Fig.~\ref{fig:6}~(a). The dashed line shows the prediction from Eqs.~\eqref{eq:rho_0_g} and \eqref{eq:radius_g} based on the SG fits of Eq.~\eqref{eq:superGauss} and the corresponding shape paraneters - see appendix~\ref{app:shape-params}. The oscillation frequency of the core can be extracted from the peak of the temporal Fourier transform of the evolution of the central density in time; this is shown by the data points in Fig.~\ref{fig:6}(b). The corresponding theoretical prediction obtained via Eq.~\eqref{eq:gamma_freq} (dotted line), again using the shape parameters from the SG profile (see appendix \ref{sec:appendix-b}), is also plotted. 

The agreement between the numerically extracted $r_c-\rho_c$ and $\omega-\rho_c$ relations and the theoretical predictions, demonstrate that the cores in our numerical simulations are indeed well captured by the corresponding properties of isolated solitons,  as approximated by the SG profile, within the error bars. This validates the use of \eqref{eq:superGauss} for describing the cores (now incorporating both the PO mode and excitations in the central region) even within more extended halos.

\begin{figure*}
    \centering
	\includegraphics[width=1.8\columnwidth]{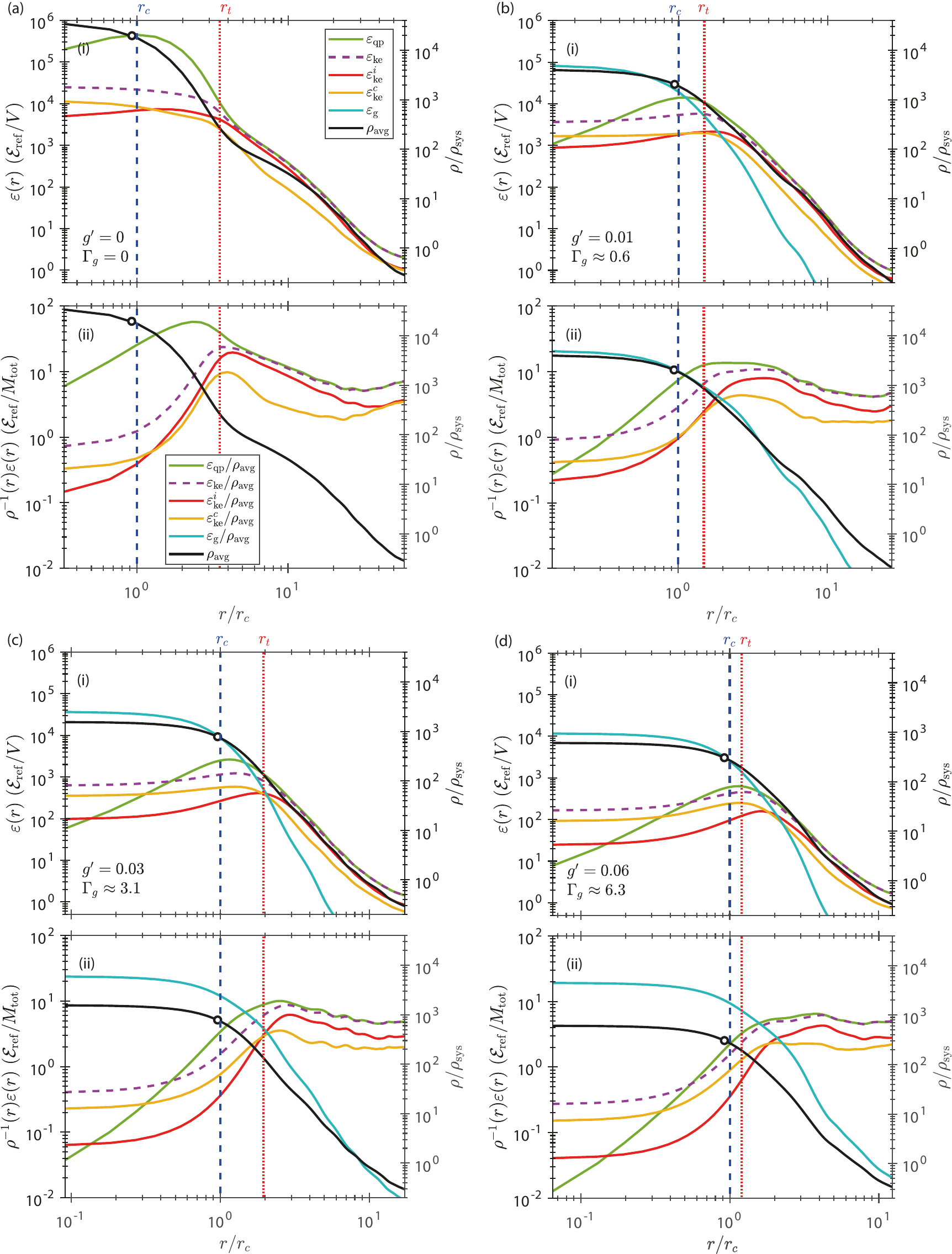}
    \caption{
    The energy distribution as a function of scaled radius $r/r_c$ for the quantum pressure energy, $\varepsilon_\mathrm{qp}$ (green), and classical kinetic energy $\varepsilon_\mathrm{ke}$ (dashed purple), decomposed into incompressible $\varepsilon_\mathrm{ke}^i$ (red) and compressible $\varepsilon_\mathrm{ke}^c$ (orange), and self interaction energy $\varepsilon_\mathrm{s}$ (cyan) components with $\rho_\mathrm{avg}$ as a reference for the data presented in Fig.~\ref{fig:3} for (a) $g^\prime=0$, (b) $10^{-2}$, (c) $3\times10^{-2}$ and (i) $6\times10^{-2}$. The circles mark the interpolated location of the half peak density, while the blue dashed lines are obtained from the SG+NFW fit.}
    \label{fig:7}
\end{figure*}

\section{Energy distribution and the Length Scales of Granules and Vortices}
\label{sec:energy}

Next, we examine the role of interactions on the distribution of compressible and incompressible classical and quantum kinetic energies and the halo's tangled vortex network, revisiting our earlier findings of (non-interacting) FDM \citep{Liu2023}.

\subsection{Energy distribution in the halo}

Such decomposed kinetic energy contributions are shown in Fig.~\ref{fig:7}~(a)-(d) in a set of 4 double-panels, for our previously considered set of interaction values of  Fig.~\ref{fig:4}, showcasing both total energies (top panels, (i)) and energies per density (bottom panels, (ii)). For clarity, such plots indicate the total angularly-averaged and time-averaged density and the relevant radial scales of the soliton, i.e.~$r_c$ and $r_t$.

In the absence of interactions [Fig.~\ref{fig:7}~(a)(i)-(ii)], we have previously demonstrated \citep{Liu2023} that the energy within the solitonic core is clearly dominated by the quantum pressure term (green lines in Fig.~\ref{fig:7}), with such term becoming comparable to the total classical kinetic energy (dashed purple lines), i.e.~the sum of compressible and incompressible kinetic energies, beyond the effective crossover region at $r_t$. Such equipartition between total classical and quantum kinetic energy components outside an FDM soliton was first noted in \citep{Mocz2017}.
We further analysed \citep{Liu2023} the components of the classical kinetic energy, finding that within $r_c$, the classical kinetic energy is clearly dominated by the compressible component (orange), while outside ($r>r_t$) the incompressible component (red) becomes more dominant. This is consistent with the picture of a vortex tangle in the outer halo, with vortices being suppressed within the soliton, while at the same time the soliton contains significant amount of sound waves -- potentially a relic of the merger scheme used to generate our soliton.
Such behaviour is clearly evident both in the total component energy plots (panels (i)) and the corresponding plots scaled to the local system density (panels (ii)).
The latter plots show that both classical and quantum kinetic energies increase with increasing radius up to $\approx r_t$, after which they start decreasing.

The remaining 3 double-panel plots [Fig.~\ref{fig:7}~(b)-(d)] examine how this picture is modified by the addition of repulsive self-interactions; now such plots also show the non-zero interaction energy (blue lines). As expected, the interaction energy decreases monotonically with increasing radius, closely following the behaviour of the system density. In practice, this distinguishes two different regions: the decrease in interaction energy is small within the solitonic core, and significant as the system transitions to the NFW outer halo profile. With increasing interaction, the system density transitions towards the well-known Thomas-Fermi profile~\citep{Bohmer2007,Chavanis2011,Harko2012,Magana2012,Abdullin:2021sro},
and such transition becomes more evident for $\Gamma_g \gtrsim 3$, were the inner core profile becomes practically flat: this observation is
consistent with the findings of \citep{Indjin2024}. Consistent with the `anomalous' behaviour of the depletion around $\Gamma_g \approx 3$, a closer comparison of the interaction energy in Fig.~\ref{fig:7} subplots (b)(ii) and (c)(ii) reveals a very slight increase of interaction energy per particle between the cases $\Gamma_g \approx 0.6$ and $\Gamma_g \approx 3.1$, again presumably associated with the previously noted change in the shape of the profile of the soliton density \citep{Indjin2024}.

We now return to the kinetic energy components, discussing the main features within the solitonic core and in the outer halo regions separately:
Within the solitonic core the dominant effect of interactions
is associated with a decrease in both the classical and quantum kinetic energies but with the quantum kinetic energy dropping the most and becoming significantly more subdominant to the classical kinetic energy at the core's centre. Indeed, in the moderately interacting regime $\Gamma_g \gtrsim 3$ its value is already less than either of compressible and incompressible kinetic energies, a trend that becomes clearer with increasing interactions. Both the compressible and incompressible kinetic energies drop less comparatively to $\varepsilon_\mathrm{qp}$, with the compressible kinetic energy becoming more dominant over the incompressible one.

Moving to the outer halo, we reach two interesting conclusions: Firstly, in all cases considered here (even for moderate strong interactions $\Gamma_g \approx 6.3$) all components of the kinetic energy (classical compressible, classical incompressible and quantum) become individually larger than the interaction energy for distances $r \gtrsim r_t$. This will have important consequences in our analysis of vortices discussed in the next subsection.
Secondly, such comparably small role of interactions in the outer halo region implies that interactions do not affect the expected equipartition between classical and quantum kinetic energy components in the outer halo \citep{Mocz2017, Hui2021}. The incompressible kinetic energy dominates over the compressible one as a relative component of the total classical kinetic energy, but the relative ratio of $E^i_\mathrm{ke}/E^c_\mathrm{ke}$ outside the core drops with increasing $g$.

Summarizing, our main finding is that interactions dominate within the (broader) solitonic region ($r \lesssim r_t$), but play little role in the outer halo. Such drastic change in the system's behaviour with increasing interactions (much beyond the $\Gamma_g \sim 1$ regime) is clearly visible by contrasting the behaviour of the non-interacting case [Fig.~\ref{fig:7}(a)], to the largest interaction strength probed here [Fig.~\ref{fig:7}(d)].

\subsection{Granule and Vortex spectra}

Building on the works of \cite{Mocz2017} and \cite{Hui2021}, our previous work \citep{Liu2023} revealed the existence of a turbulent structure of vortices in the outer FDM halo, consistent with the significant incompressible kinetic energy in such region: such vortex tangle evolved dynamically, with vortices moving on relatively slow cosmological timescales, but revealed no evidence of decay, as relevant in interacting laboratory-based superfluids (see, e.g.~\citep{2002PhRvA..66a3603B,Baggaley2012}).
In the latter systems, involving local interactions, decay of the vortex tangle is expected to arise through mode mixing facilitated by the nonlocal interaction term, resulting in an energy cascade transferring energy from the incompressible to the compressible sector. A natural arising question we address here is whether the addition of local self-interactions, modifying the SPE to the GPPE system of equations would be enough to cause a similar energy cascade and a vortex tangle decay.

\begin{figure}
	\includegraphics[width=1\columnwidth]{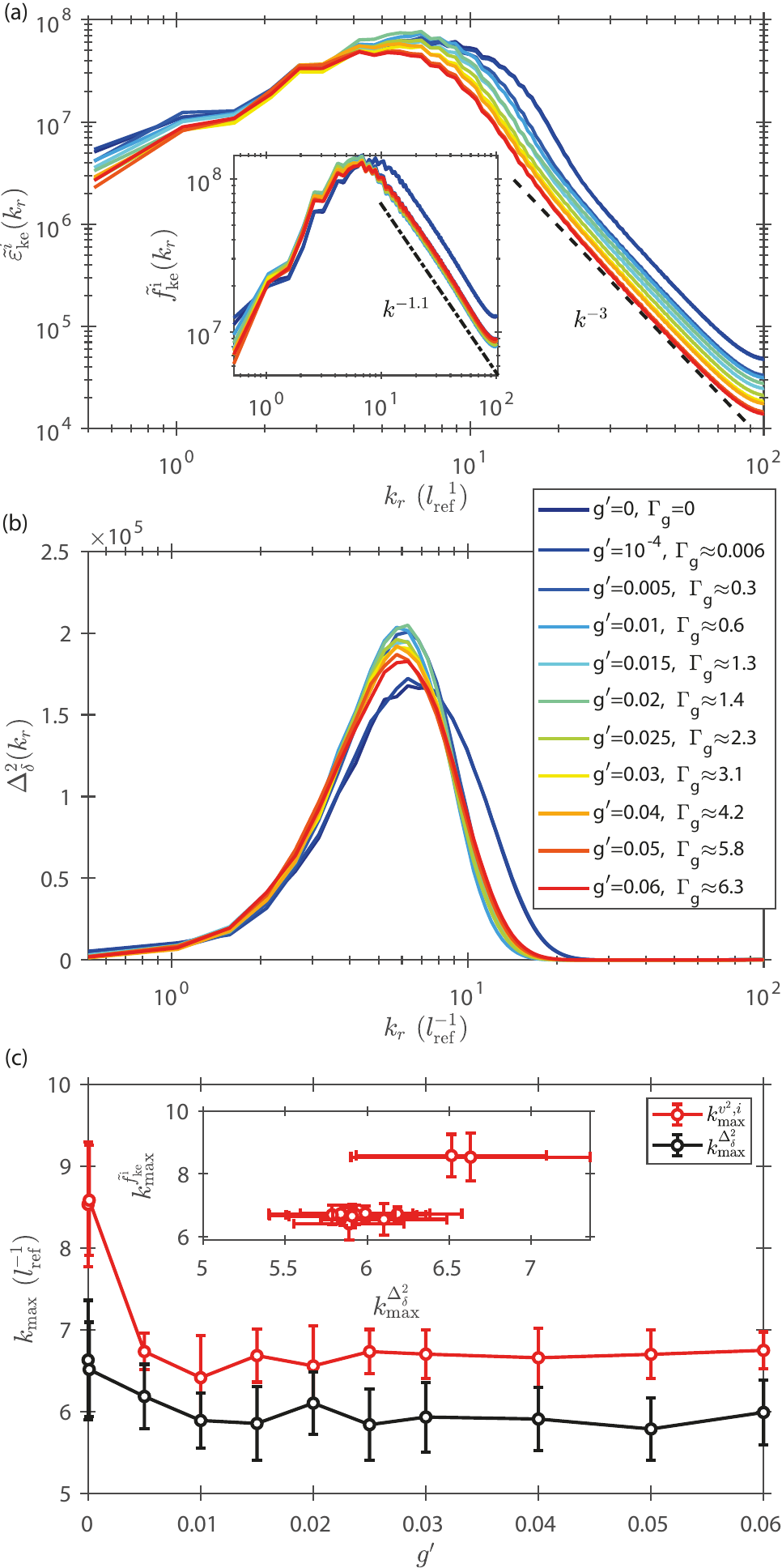}
    \caption{ (a) Incompressible kinetic energy and (b) overdensity power Spectra and (c) their peak location (bottom), as functions of $g^\prime$. The inset in (a) shows the incompressible velocity spectra, and that in (c) shows the relation between the peak locations of the overdensity power and incompressible velocity spectra. It shows that the spectra do not dependent on much on the self interaction when $\Gamma_g\gtrsim\mathcal{O}(1)$.
    }
    \label{fig:8}
\end{figure}

The key first quantity thus to examine is the spectrum of the incompressible kinetic energy: this is useful as it coincides with the vortex energy spectrum \citep{1997PhFl....9.2644N, 1997PhRvL..78.3896N,2012arXiv1202.1863T,2012PhRvB..86j4501B,2012PhRvL.109t5304B}. 
We compute this in the form of the cumulative measure of the incompressible energy density amplitude's Fourier components via~\citep{1997PhFl....9.2644N} 
\be
\tilde{\epsilon}_\mathrm{ke}^{i}(k_r)\equiv\int d\mathbf{\Omega}_k\,k^2\tilde{\epsilon}_\mathrm{ke}^{i}(\mb{k})=\sum_{k_r\leq|\mb{k}|\leq k_r+\Delta k}\tilde{\epsilon}_\mathrm{ke}^{i}(\mb{k})
\ee
limiting the range to $k_r\leq|\mb{k}|<k_r+\Delta k$, where $\mathbf{\Omega}_k$ is the solid angle in the momentum space and $\Delta k = 2 \pi/L$ with $L=12L_\mathrm{ref}$.
The behaviour of $\tilde{\epsilon}_\mathrm{ke}^{i}(k_r)$ vs. $k_r$ is shown in Fig.~\ref{fig:8}(a) for different interaction strengths, with the blue line corresponding to the $g=0$ case previously discussed in~\citep{Liu2023}. The addition of interactions leads to small shift of the position of the peak to larger scales, with a corresponding decrease in the area under the graph and hence the value of the integrated incompressible kinetic energy $E_{\rm kin}^i$, consistent with fig.~\ref{fig:7}. We note there is no detectable change in the \emph{slope} of the spectrum beyond the peak, 
which indicates clear evidence of the existence of well-formed quantum vortex cores. Moreover, as already mentioned in~\citep{Liu2023}, a clear $k^{-3}$ tail for large $k$ unveils the quantum vortex core structure where $\rho(r)v(r)\approx\mathrm{constant}$ around vortex cores.
However, the lack of other prevailing scaling behaviours beyond the peak, such as an extended $k$ regime displaying the conventional Kolmogorov turbulence, suggests the presence of disorder in the vortex tangles and hints at the existence of disorganised quantum (or ``Vinen'' or ``ultraquantum'') turbulence~\citep{Baggaley2012}\footnote{We thank Carlo Barenghi for clarifying the relevant terminology.}.

A complementary quantity for characterising the vortex tangle spectrum is the incompressible velocity spectrum
\be
\tilde{f}_\mathrm{ke}^\mathrm{i}=\int d\mathbf{\Omega}_kk^2\int d\mathbf{r}e^{-i\mathbf{k}\cdot\mathbf{r}}\frac{|\mathbf{v}^\mathrm{i}(\mb{r})|^2}{2}\,.
\ee
For a standard quantum vortex, the spectrum is expected to exhibit $k^{-1}$ at large momenta, reflecting the $v^\mathrm{i}\propto1/r$ behaviour around the vortex core~\citep{1997PhFl....9.2644N}. As shown in the inset of Fig.~\ref{fig:8}, we find the scaling at large momentum being $\approx k^{-1.1}$, consistent with what has been found in the full velocity spectrum~\citep{Mocz2017}, and suggesting that $v^\mathrm{i}\propto r^{-0.9}$ with the density around the vortex core being $\propto r^{1.8}$, recovering the standard vortex energy spectrum that $\rho |\mathbf{v}^\mathrm{i}|^2\approx \mathrm{const.}$. 
The peak location of incompressible velocity spectrum is consistent with the peak location of the  $\tilde{\epsilon}_\mathrm{ke}^{i}(k_r)$ spectrum. The peaks in these spectra provide a characteristic scale that characterises the inter-vortex distance in the halo.

To investigate the interplay of vortex cores and the granules in the FDM literature, we consider the power spectrum of the overdensity
\be
\delta\rho(\mb{r},t)=\frac{\rho(\mb{r},t)-\rho_{\rm avg}(\mathbf{r})}{\bar{\rho}(\mathbf{r})} \;,
\label{eq:density_fluc}
\ee
where $\rho_{\rm avg}(\mathbf{r})$ is the spherically symmetric density profile obtained by time averaging. Its power spectrum is defined as~\citep{Liu2023}
\ba{rl}
\Delta^2_\delta(k_r,t)=&\displaystyle\frac{k_r^3}{(2\pi^2)4\pi k_r^2}\int d\mb{\Omega}_kP_\delta(\mb{k},t)
\\\\
\approx&\displaystyle\frac{k_r^3}{2\pi^2 \mathcal{N}_{k_r}}\sum_{k_r\leq|\mb{k}|<k_r+\Delta k}|\mb{k}|P_\delta(\mb{k},t)
\ea
where $P_\delta(\mb{k},t)=|\tilde{\eta}_\delta(\mb{k},t)|^2$ and $\tilde{\eta}_\delta(\mb{k},t)$ is the Fourier transformation of the overdensity $\delta \rho(\mb{r},t)$.
The time averaged power spectrum of the overdensity is shown in Fig.~\ref{fig:8}(b). The previously-studied $g=0$ spectrum (dark blue) shifts its peak to a somewhat smaller wavenumber as soon as a non-negligible non-zero interaction is introduced (as visible, e.g., already for $\Gamma_g \approx 0.3 $), also increasing somewhat in amplitude.
    Beyond such value, and for the entire range of weak to moderate interactions ($\Gamma_g \lesssim 6.3$) probed in this work, for which $(g \rho_c)$ remains approximately constant, the position of the maximum of the overdensity spectrum appears to remain at largely the same wavenumber with its amplitude gradually decreasing with $g$.

The relation between the peak locations of the incompressible velocity spectrum, $k_\mathrm{max}^{\tilde{f}^\mathrm{i}_\mathrm{ke}}$, and the overdensity power spectrum, $k_\mathrm{max}^{\Delta^2_\delta}$ is shown in the inset of  Fig.~\ref{fig:8}~(c).
This shows that the previously obtained result~\citep{Liu2023}
$k_\mathrm{max}^{|\mathbf{v}^i|^2} \approx 0.93 k_\mathrm{max}^{\Delta^2_\delta}$  obtained via the SPE, remains also approximately valid in the presence of interactions: note that the two `outlier' points correspond to the $g \approx 0$ limit studied previously.
The dependence of such $k_\mathrm{max}$ on interaction strength, is best visualised in the main plot of Fig.~\ref{fig:8}(c). Within our numerical error bars, it becomes evident that both independently determined $k_\mathrm{max}$ are approximately independent of interactions for $\Gamma_g \gtrsim 0.3$.
Importantly, this shows that the similarity of the mean inter-vortex distance and mean granule scale, known to hold in the $g=0$ case  is largely unaffected by the repulsive interactions. Hence, despite $g\neq 0$ these two scales are still related to the local de Broglie wavelength in our simulations, as expected \citep{Mocz2017,2021JCAP...01..011H, Schobesberger:2021ghi}.

While this may on first inspection appear somewhat counterintuitive from a condensed matter perspective -- as the vortex core size of a gas interacting via weak contact interactions is in fact fixed by the interaction strength -- such results should already have been anticipated by the observation that the interaction region in the outer halo is in fact negligible, at least compared to all individual kinetic and quantum energy components [see e.g. Fig.~\ref{fig:7}(d)(ii)].

\begin{figure}
	\includegraphics[width=1\columnwidth]{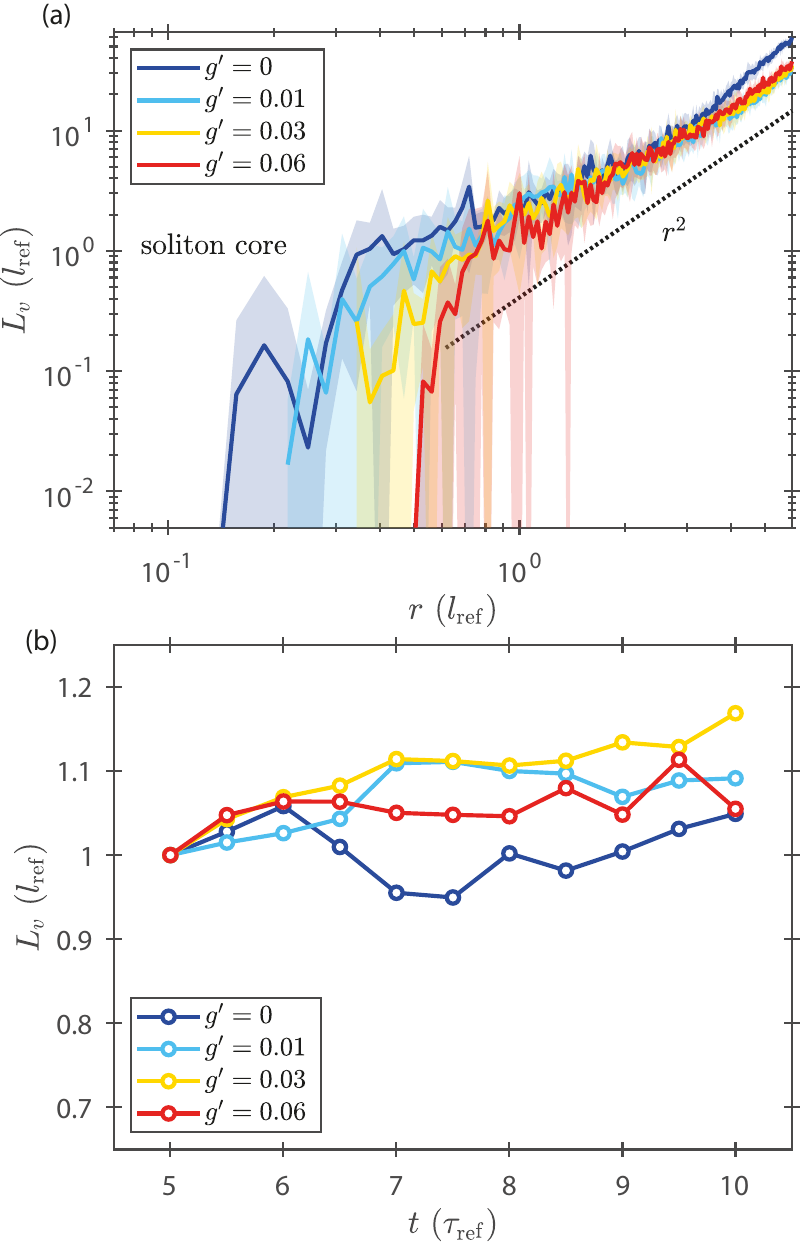}
    \caption{(a) Shell vortex line length, $L_v$ as a function of radius. Outside the soliton core, i.e.~for $r>r_c(g')$, we see the emergence of an $L_v \sim r^2$ scaling, implying the shell vortex density is constant. (b) The evolution of total vortex line length scaled by its value at $t=5\tau_\mathrm{ref}$ shows there is no evidence of vortex decaying and significant dependence on self interaction strength.}
    \label{fig:9}
\end{figure}

\subsection{Vortex linelength}

Turbulence is known to decay in laboratory-based quantum fluids~\citep{1997PhFl....9.2644N,2002PhRvA..66a3603B,Bradley2006}. This can be attributed to phonon emissions via Kelvin waves~\citep{Leadbeater2002,Leadbeater2003}, or vortex reconnections~\citep{Leadbeater2001,Zuccher2012,Baggaley2012}, with an additional channel provided by the interaction with the thermal component~\citep{1997PhFl....9.2644N}. However, the evolution of energy in the context of our examined SFDM halos (Fig.~\ref{fig:3}), does not show any signs of decrease in the incompressible kinetic energy. To corroborate this observation and further characterise turbulent behaviour in outer halos, we perform a direct investigation of the total vortex line length and shell density. Specifically, the vortex cores are detected on the subgrid level by interpolating $\Psi$, probing the $2\pi$ phase winding and density minimum. Then the vortex points are linked via the least distance algorithm with a threshold $\Delta x/8\approx0.004L_\mathrm{ref}$ to estimate the total vortex line length.

The cumulative integrated vortex line length $L_v$ up to radius $r$ is shown in Fig.~\ref{fig:9}~(a) as a function of $r$.
A clear transition from a negligible vortex line length (within the solitonic core) to an $L_v \sim r^2$ dependence of the detected length scale is found for all interaction values probed; this indicates that mean vortex densities on each shell are constant, with no discernible dependence on the strength of self interaction. The one  effect of $g\neq 0$ in the weak-to-moderate interaction regime considered here is to increase the outer solitonic radius $r_t$ (as already seen in Fig.~\ref{fig:1}) and thus push the region where vorticity starts emerging  to larger $r$ values: this is clearly visible by the shift of the curves from the left to the right.

To directly address the question of the stability (or potential decay) of the arising vortex tangle, we also look at the evolution of the total vortex line length as a function of time, focusing only on late timescales after virialisation has been achieved (as evidenced by the near constant interaction energies in Fig.~\ref{fig:3}). The approximately constant (within our numerical determination capabilities) value of the vortex line length with time irrespective of the interaction strength indeed confirms that the vortex tangle does not noticeably decay over the probed cosmologically relevant timescale, for the moderately strong -- yet observationally relevant~\citep{Indjin:2025lqs} -- interaction strengths probed here.

\section{Discussion and Conclusions} \label{sec:conclusions}

In this work we have studied the role of boson repulsive self-interactions on the static and dynamic properties of isolated Scalar Field Dark Matter cored halos, with emphasis on coherence properties, condensate fraction, core oscillations and quantised vorticity, thereby extending our earlier $g=0$ FDM findings \citep{Liu2023} on such matters. The present study is limited to the weak to moderate (repulsive) interaction regime $0 \le g/g_* \lesssim 6.3$, where $g_*$ [Eq.~(\ref{eq:g_*})] is an indicative characteristic interaction strength at which the interaction energy of the core equals its quantum kinetic energy. Our previous work \citep{Indjin2024} has characterised the shape of the profile transition from the usual $g=0$ FDM empirical profile \citep{Schive2014,Schive2014a,Mocz2017} to the strongly-interacting Thomas-Fermi one~\citep{Bohmer2007,Chavanis2011,Harko2012,Magana2012,Abdullin:2021sro} finding that this occurs around a dimensionless interaction strength $\Gamma_g = g/g_* \approx 3$ (see Fig.~3 of \citep{Indjin2024}). 

Our parallel study \citep{Indjin:2025lqs} analysing rotation curve data from the SPARC galaxy dataset ~\citep{SPARC} has also shown that: (i) such a transition can be well characterised by a Super-Gaussian profile [Eq.~(\ref{eq:superGauss})], used throughout this work, 
and that (ii) the probed dark-matter-dominated SPARC galaxies appear to be consistently better fit by a distinctly non-zero interaction strength\footnote{More specifically, our analysis of SPARC galaxy datasets indicates a single $(m,g)$ point in the space of boson masses $m$ and self-coupling constants $g$:
$\log_{10}\left(m \,[\mathrm{eV}/c^2] \right) = \log_{10}(1.98)-22^{+0.8}_{-0.6}$ and $\log_{10}\left(g \, [\mathrm{Jm}^3/kg] \right) = \log_{10}(1.45)-28^{+0.4}_{-1.2}$.}
consistent with earlier findings~\citep{Delgado2022,Chakrabarti2022}. While such values remain to be confirmed by broader galactic studies and span a broad range of dimensionless interaction strengths $4.8 \lesssim g/g_* \lesssim 1630$, a key emerging point of that analysis relevant to the present discussion is that there appear to exist galaxies whose rotation curves are well described by dimensionless interaction strengths $g/g_*$ within the range probed in the present work. For example, we note here the galaxy KK98-251, which was found \citep{Indjin:2025lqs} to have an inferred value of $g / g_* \approx 4.8$. 

The key findings of the present work for the probed regime are as follows: Given an initial spatial configuration of 10 density concentrations with Gaussian profiles and with similar masses,  the main effect of increasing repulsive interactions is to expand the resulting solitons and reduce their central density, with interaction energy replacing the quantum pressure in acting as the main factor preventing the soliton from collapsing under its own gravity. For the parameter values investigated here, the most strongly interacting soliton cores of the resulting halos intrude into the Thomas-Fermi regime but remain moderately interacting, with the maximum achieved $g/g_* \equiv \Gamma_g \approx 6.3$ and with mildly increasing mass - see Fig.~\ref{fig:2} and table 1. An increasing $g$, resulting in a higher overall total energy in the system initialization, leads to the formation of  solitonic cores with lower central density and a broader size, resulting in an effective shallower gravitational potential and non-negligible condensate depletion. 
Our discussion of depletion was underpinned by the fact that the numerical study of the Gross-Pitaevskii-Poisson equations facilitates a simultaneous analysis of both coherent and incoherent modes of the system, noting that the condensate should only be identified as the coherent part of the entire simulated classical field, in the sense of the Penrose-Onsager mode corresponding to the largest eigenvalue.

Due to the significantly decreasing dark matter density with radius, the self-interaction term is subdominant in the outer halo regions and hence the role of interactions remains rather limited there: We still find the existence of slowly-evolving quantum vortices comprising a self-sustaining dynamical vortex tangle, which does not however appear to decay over probed timescales, and density fluctuations (granules) whose typical length scale is very similar to the inter-vortex distance. Interactions change these scales somewhat but to a far smaller degree relative to the expansion and dilution of the core with increasing $g$. Hence, the characteristic scale of our outer halos is still the de Broglie wavelength as in the $g=0$ FDM case, while the cores tend towards the Thomas-Fermi regime\footnote{We are grateful to the anonymous reviewer for pointing the relevance of these two different scales in the core and the halo.}, with some of them having already reached their Thomas-Fermi radius (see Fig.~\ref{fig:2}). These findings are consistent with the picture of SFDM cored halos put forward in \citep{Dawoodbhoy2021} of Thomas-Fermi cores surrounded by enveloping halos supported by the effects of quantum pressure and classical kinetic energy (which we find in equipartition, as in FDM). 
 
Furthermore, we speculated on a possible explanation of the observed increasing condensate depletion with interaction strength. Using an {\em indirect} analogy to harmonically-confined cold atomic gases, we noted that the gravitational potential of the steady-state solitonic core becomes softer with increasing interactions. Treating the gravitational potential as approximately harmonic in the centre, this softening implies a reduction in the effective energy level spacing, and thus an increased relative role of interaction energy when compared to such level spacing. This makes more excited modes accessible for a given configuration thus enhnancing the depletion. Of course a more quantitative study of the mode spectrum would be required to verify and quantify this hypothesis.

The solitons that formed in the core of the halos exhibited oscillations of their central density. The dependence of their frequencies on $g$ was found to be well described by results for isolated, ideal solitons from \citep{Indjin2024}. The emergence of (at least mildly) excited solitons in the case of $g\neq 0$ merger simulations was also receently pointed out in~\citep{RindlerDaller2025}.
A full understanding of the final dynamical solitonic steady-state solution across a broader interaction strength range, and the relation between solitonic random walks \citep{Schive2020PhRvL.124t1301S,Li2020,DuttaChowdhury2021,Zagorac2022}, oscillations and excitations, and depletion -- including the relation between the Penrose-Onsager mode and the ground state mode in decompositions -- see \citep{Lin2018,Yavetz2021,Zagorac2022,Zagorac2023,Chan2023,Zimmermann2025} for the SPE and \citep{Salasnich2025} for the GPPE -- are deferred to future work.

The existence of vortices (outside the solitonic core, as found in our simulations) may in fact provide future observational signatures, as they characterise the granule distribution, whose existence may be detectable in future gravitational lensing obervations~\citep{Chan2020-lensing,Laroche2022-lensing,Powell2023-lensing,Amruth2023-lensing,Vegetti2024-lensing} (or other observations), extending much beyond present capabilities. It is also worth noting the recent works \citep{Korshynska_2023,Korshynska_2025,Asakawa_2024,Brax:2025uaw, Brax:2025vdh} where solitons containing vortices in an SFDM case are found - see also \citep{Rindler-Daller:2011afd} for earlier work. In contrast, our simulations create vortex-free solitons, which expel any vortices that might approach from the surrounding halo, {as also happens in the non-interacting FDM case \citep{Schobesberger:2021ghi,Liu2023}. Comparisons between these different regimes are clearly worthy of further numerical investigation.              

Finally, we point out that  the prevailing bimodal picture of a central solitonic condensate surrounded by an incoherent NFW profile, combined with our presented 
evidence for depletion within the central solitonic core and the potential for self-sustaining quantum vorticity in the outer regions, raises significant questions which can be further addressed within the context of a {\em hybrid} interacting Fuzzy Dark Matter and Cold Dark Matter-based model.
Such a model has been formally developed in \citep{Proukakis:2023nmm, Proukakis:PhysRevD.111.023505},
and further discussed in~\citep{Proukakis:2023txk}.
Investigation of such a generalised hybrid model, might also be relevant in relation to
condensation relaxation times within the Fuzzy Dark Matter models~\citep{Levkov2018,chanda_formation_early,chanda_formation,marsh_formation}.

\section*{Authorship statement}
This work was conceived by IKL, NPP and GR; numerical simulations were conducted and analysed by IKL, in coordination with MI, with figures plotted under the guidance of NPP and GR; the vortex analysis of our SFDM data, leading to the data shown in Fig.~\ref{fig:9}, was done by NK, in discussions with IKL who plotted the final figures. The paper was written by IKL, NPP and GR with critical input from MI.

\section*{Acknowledgements}

IKL acknowledges funding from the Marie Sklodowska- Curie grant agreement No. 897324 (upgradeFDM) and the Science and Technology Facilities Council (grant: ST/W001020/1), while GR and NPP acknowledge funding from the Leverhulme Trust (Grant no. RPG-2021-010). We also acknowledge discussions with Alex Soto and Carlo Barenghi. The simulations were conducted on the HPC Cluster Rocket of Newcastle University. We wish to thank the anonymous reviewer for many pertinent points which allowed us to significantly improve the manuscript.

\section*{Data Availability}
Data supporting this publication can be openly accessed under an `Open Data Commons Open Database License' [...to appear on publication...]

\bibliographystyle{mnras}

\begin{thebibliography}{}
\makeatletter
\relax
\def\mn@urlcharsother{\let\do\@makeother \do\$\do\&\do\#\do\^\do\_\do\%\do\~}
\def\mn@doi{\begingroup\mn@urlcharsother \@ifnextchar [ {\mn@doi@}
  {\mn@doi@[]}}
\def\mn@doi@[#1]#2{\def\@tempa{#1}\ifx\@tempa\@empty \href
  {http://dx.doi.org/#2} {doi:#2}\else \href {http://dx.doi.org/#2} {#1}\fi
  \endgroup}
\def\mn@eprint#1#2{\mn@eprint@#1:#2::\@nil}
\def\mn@eprint@arXiv#1{\href {http://arxiv.org/abs/#1} {{\tt arXiv:#1}}}
\def\mn@eprint@dblp#1{\href {http://dblp.uni-trier.de/rec/bibtex/#1.xml}
  {dblp:#1}}
\def\mn@eprint@#1:#2:#3:#4\@nil{\def\@tempa {#1}\def\@tempb {#2}\def\@tempc
  {#3}\ifx \@tempc \@empty \let \@tempc \@tempb \let \@tempb \@tempa \fi \ifx
  \@tempb \@empty \def\@tempb {arXiv}\fi \@ifundefined
  {mn@eprint@\@tempb}{\@tempb:\@tempc}{\expandafter \expandafter \csname
  mn@eprint@\@tempb\endcsname \expandafter{\@tempc}}}

\bibitem[\protect\citeauthoryear{Abdullin \& Popov}{Abdullin \&
  Popov}{2021}]{Abdullin:2021sro}
Abdullin I.~G.,  Popov V.~A.,  2021, \mn@doi [JCAP]
  {10.1088/1475-7516/2021/11/055}, 11, 055

\bibitem[\protect\citeauthoryear{Al~Khawaja, Andersen, Proukakis  \&
  Stoof}{Al~Khawaja et~al.}{2002}]{PhysRevA.66.013615}
Al~Khawaja U.,  Andersen J.~O.,  Proukakis N.~P.,   Stoof H. T.~C.,  2002,
  \mn@doi [Phys. Rev. A] {10.1103/PhysRevA.66.013615}, 66, 013615

\bibitem[\protect\citeauthoryear{Alvarez-R{\'{i}}os, Guzm{\'{a}}n  \&
  Shapiro}{Alvarez-R{\'{i}}os et~al.}{2023}]{Alvarez-Rios2023}
Alvarez-R{\'{i}}os I.,  Guzm{\'{a}}n F.~S.,   Shapiro P.~R.,  2023, \mn@doi
  [Physical Review D] {10.1103/PhysRevD.107.123524}, 107, 123524

\bibitem[\protect\citeauthoryear{Amruth et~al.,}{Amruth
  et~al.}{2023}]{Amruth2023-lensing}
Amruth A.,  et~al., 2023, \mn@doi [Nature Astronomy]
  {10.1038/s41550-023-01943-9}, 7, 736

\bibitem[\protect\citeauthoryear{Asakawa \& Tsubota}{Asakawa \&
  Tsubota}{2024}]{Asakawa_2024}
Asakawa K.,  Tsubota M.,  2024, \mn@doi [Physical Review A]
  {10.1103/physreva.110.053310}, 110

\bibitem[\protect\citeauthoryear{Ba\~nares Hern\'andez, Castillo,
  Martin~Camalich  \& Iorio}{Ba\~nares Hern\'andez
  et~al.}{2023}]{Banares-Hernandez2023b}
Ba\~nares Hern\'andez A.,  Castillo A.,  Martin~Camalich J.,   Iorio G.,  2023,
  \mn@doi [Astron. Astrophys.] {10.1051/0004-6361/202346686}, 676, A63

\bibitem[\protect\citeauthoryear{{Baggaley}, {Barenghi}  \&
  {Sergeev}}{{Baggaley} et~al.}{2012a}]{Baggaley2012}
{Baggaley} A.~W.,  {Barenghi} C.~F.,   {Sergeev} Y.~A.,  2012a, \mn@doi [\prb]
  {10.1103/PhysRevB.85.060501}, \href
  {https://ui.adsabs.harvard.edu/abs/2012PhRvB..85f0501B} {85, 060501}

\bibitem[\protect\citeauthoryear{{Baggaley}, {Sherwin}, {Barenghi}  \&
  {Sergeev}}{{Baggaley} et~al.}{2012b}]{2012PhRvB..86j4501B}
{Baggaley} A.~W.,  {Sherwin} L.~K.,  {Barenghi} C.~F.,   {Sergeev} Y.~A.,
  2012b, \mn@doi [\prb] {10.1103/PhysRevB.86.104501}, \href
  {https://ui.adsabs.harvard.edu/abs/2012PhRvB..86j4501B} {86, 104501}

\bibitem[\protect\citeauthoryear{{Baggaley}, {Laurie}  \&
  {Barenghi}}{{Baggaley} et~al.}{2012c}]{2012PhRvL.109t5304B}
{Baggaley} A.~W.,  {Laurie} J.,   {Barenghi} C.~F.,  2012c, \mn@doi [\prl]
  {10.1103/PhysRevLett.109.205304}, \href
  {https://ui.adsabs.harvard.edu/abs/2012PhRvL.109t5304B} {109, 205304}

\bibitem[\protect\citeauthoryear{Baldeschi, Ruffini  \& Gelmini}{Baldeschi
  et~al.}{1983}]{Baldeschi:1983mq}
Baldeschi M.~R.,  Ruffini R.,   Gelmini G.~B.,  1983, \mn@doi [Phys. Lett. B]
  {10.1016/0370-2693(83)90688-3}, 122, 221

\bibitem[\protect\citeauthoryear{Barenghi, Skrbek  \& Sreenivasan}{Barenghi
  et~al.}{2014}]{cfb-review}
Barenghi C.~F.,  Skrbek L.,   Sreenivasan K.~R.,  2014, \mn@doi [Proceedings of
  the National Academy of Sciences] {10.1073/pnas.1400033111}, 111, 4647

\bibitem[\protect\citeauthoryear{Berezhiani, Cintia, De~Luca  \&
  Khoury}{Berezhiani et~al.}{2025}]{Berezhiani:2025maf}
Berezhiani L.,  Cintia G.,  De~Luca V.,   Khoury J.,  2025, 2505.23900
  [astro-ph.Co]

\bibitem[\protect\citeauthoryear{{Berloff} \& {Svistunov}}{{Berloff} \&
  {Svistunov}}{2002}]{2002PhRvA..66a3603B}
{Berloff} N.~G.,  {Svistunov} B.~V.,  2002, \mn@doi [\pra]
  {10.1103/PhysRevA.66.013603}, \href
  {https://ui.adsabs.harvard.edu/abs/2002PhRvA..66a3603B} {66, 013603}

\bibitem[\protect\citeauthoryear{Bernal, Fern\'andez-Hern\'andez, Matos  \&
  Rodr\'\i{}guez-Meza}{Bernal et~al.}{2018}]{Bernal2017}
Bernal T.,  Fern\'andez-Hern\'andez L.~M.,  Matos T.,   Rodr\'\i{}guez-Meza
  M.~A.,  2018, \mn@doi [Mon. Not. Roy. Astron. Soc.] {10.1093/mnras/stx3208},
  475, 1447

\bibitem[\protect\citeauthoryear{Binney \& Tremaine}{Binney \&
  Tremaine}{2008}]{binney2008galactic}
Binney J.,  Tremaine S.,  2008, Galactic Dynamics: Second Edition.
Princeton Series in Astrophysics, Princeton University Press, \url
  {https://books.google.co.uk/books?id=qxWt20TH--cC}

\bibitem[\protect\citeauthoryear{{Blakie} \& {Davis}}{{Blakie} \&
  {Davis}}{2005}]{2005PhRvA..72f3608B}
{Blakie} P.~B.,  {Davis} M.~J.,  2005, \mn@doi [\pra]
  {10.1103/PhysRevA.72.063608}, \href
  {https://ui.adsabs.harvard.edu/abs/2005PhRvA..72f3608B} {72, 063608}

\bibitem[\protect\citeauthoryear{{Blakie}, {Bradley}, {Davis}, {Ballagh}  \&
  {Gardiner}}{{Blakie} et~al.}{2008}]{2008AdPhy..57..363B}
{Blakie} P.~B.,  {Bradley} A.~S.,  {Davis} M.~J.,  {Ballagh} R.~J.,
  {Gardiner} C.~W.,  2008, \mn@doi [Advances in Physics]
  {10.1080/00018730802564254}, \href
  {https://ui.adsabs.harvard.edu/abs/2008AdPhy..57..363B} {57, 363}

\bibitem[\protect\citeauthoryear{Blakie, Baillie  \& Bisset}{Blakie
  et~al.}{2013}]{Blakie_Depletion}
Blakie P.~B.,  Baillie D.,   Bisset R.~N.,  2013, \mn@doi [Phys. Rev. A]
  {10.1103/PhysRevA.88.013638}, 88, 013638

\bibitem[\protect\citeauthoryear{{Bradley} et~al.,}{{Bradley}
  et~al.}{2006}]{Bradley2006}
{Bradley} D.~I.,  et~al., 2006, \mn@doi [\prl] {10.1103/PhysRevLett.96.035301},
  \href {https://ui.adsabs.harvard.edu/abs/2006PhRvL..96c5301B} {96, 035301}

\bibitem[\protect\citeauthoryear{Brax \& Valageas}{Brax \&
  Valageas}{2025a}]{Brax:2025uaw}
Brax P.,  Valageas P.,  2025a, \mn@doi [Phys. Rev. D]
  {10.1103/PhysRevD.111.103527}, 111, 103527

\bibitem[\protect\citeauthoryear{Brax \& Valageas}{Brax \&
  Valageas}{2025b}]{Brax:2025vdh}
Brax P.,  Valageas P.,  2025b, \mn@doi [Phys. Rev. D] {10.1103/s91m-pldz}, 111,
  103538

\bibitem[\protect\citeauthoryear{Böhmer \& Harko}{Böhmer \&
  Harko}{2007}]{Bohmer2007}
Böhmer C.~G.,  Harko T.,  2007, \mn@doi [Journal of Cosmology and
  Astroparticle Physics] {10.1088/1475-7516/2007/06/025}, 2007, 025

\bibitem[\protect\citeauthoryear{Chakrabarti, Dave, Dutta  \&
  Goswami}{Chakrabarti et~al.}{2022}]{Chakrabarti2022}
Chakrabarti S.,  Dave B.,  Dutta K.,   Goswami G.,  2022, \mn@doi [Journal of
  Cosmology and Astroparticle Physics] {10.1088/1475-7516/2022/09/074}, 2022,
  074

\bibitem[\protect\citeauthoryear{{Chan}, {Schive}, {Woo}  \& {Chiueh}}{{Chan}
  et~al.}{2018}]{2018MNRAS.478.2686C}
{Chan} J. H.~H.,  {Schive} H.-Y.,  {Woo} T.-P.,   {Chiueh} T.,  2018, \mn@doi
  [\mnras] {10.1093/mnras/sty900}, \href
  {https://ui.adsabs.harvard.edu/abs/2018MNRAS.478.2686C} {478, 2686}

\bibitem[\protect\citeauthoryear{Chan, Schive, Wong, Chiueh  \&
  Broadhurst}{Chan et~al.}{2020}]{Chan2020-lensing}
Chan J. H.~H.,  Schive H.-Y.,  Wong S.-K.,  Chiueh T.,   Broadhurst T.,  2020,
  \mn@doi [Phys. Rev. Lett.] {10.1103/PhysRevLett.125.111102}, 125, 111102

\bibitem[\protect\citeauthoryear{Chan, Ferreira, May, Hayashi  \& Chiba}{Chan
  et~al.}{2021}]{Chan2021}
Chan H. Y.~J.,  Ferreira E. G.~M.,  May S.,  Hayashi K.,   Chiba M.,  2021, 10,
  1

\bibitem[\protect\citeauthoryear{Chan, Ferreira, May, Hayashi  \& Chiba}{Chan
  et~al.}{2022}]{Chan2022}
Chan H. Y.~J.,  Ferreira E. G.~M.,  May S.,  Hayashi K.,   Chiba M.,  2022,
  \mn@doi [Monthly Notices of the Royal Astronomical Society]
  {10.1093/MNRAS/STAC063}, 511, 943

\bibitem[\protect\citeauthoryear{Chan, Sibiryakov  \& Xue}{Chan
  et~al.}{2023}]{Chan2023}
Chan J. H.-H.,  Sibiryakov S.,   Xue W.,  2023, \mn@doi [JHEP]
  {10.1007/JHEP08(2023)045}, 08, 045

\bibitem[\protect\citeauthoryear{{Chandrasekhar}}{{Chandrasekhar}}{1939}]{1939isss.book.....C}
{Chandrasekhar} S.,  1939, {An introduction to the study of stellar structure}

\bibitem[\protect\citeauthoryear{Chang, Bouton, Cayla, Qu, Aspect, Westbrook
  \& Cl\'ement}{Chang et~al.}{2016}]{atomic-depletion}
Chang R.,  Bouton Q.,  Cayla H.,  Qu C.,  Aspect A.,  Westbrook C.~I.,
  Cl\'ement D.,  2016, \mn@doi [Phys. Rev. Lett.]
  {10.1103/PhysRevLett.117.235303}, 117, 235303

\bibitem[\protect\citeauthoryear{Chavanis}{Chavanis}{2011}]{Chavanis2011}
Chavanis P.~H.,  2011, \mn@doi [Physical Review D - Particles, Fields,
  Gravitation and Cosmology] {10.1103/PhysRevD.84.043531}, 84, 1

\bibitem[\protect\citeauthoryear{Chavanis}{Chavanis}{2021}]{PhysRevD.103.123551}
Chavanis P.-H.,  2021, \mn@doi [Phys. Rev. D] {10.1103/PhysRevD.103.123551},
  103, 123551

\bibitem[\protect\citeauthoryear{Chavanis \& Delfini}{Chavanis \&
  Delfini}{2011}]{Chavanis2011Delfini}
Chavanis P.-H.,  Delfini L.,  2011, \mn@doi [Physical Review D]
  {10.1103/PhysRevD.84.043532}, 84, 043532

\bibitem[\protect\citeauthoryear{{Chen}, {Du}, {Lentz}, {Marsh}  \&
  {Niemeyer}}{{Chen} et~al.}{2021}]{Chen2021}
{Chen} J.,  {Du} X.,  {Lentz} E.~W.,  {Marsh} D. J.~E.,   {Niemeyer} J.~C.,
  2021, \mn@doi [\prd] {10.1103/PhysRevD.104.083022}, \href
  {https://ui.adsabs.harvard.edu/abs/2021PhRvD.104h3022C} {104, 083022}

\bibitem[\protect\citeauthoryear{Chen, Du, Lentz  \& Marsh}{Chen
  et~al.}{2022}]{marsh_formation}
Chen J.,  Du X.,  Lentz E.~W.,   Marsh D. J.~E.,  2022, \mn@doi [Phys. Rev. D]
  {10.1103/PhysRevD.106.023009}, 106, 023009

\bibitem[\protect\citeauthoryear{Chiang, Schive  \& Chiueh}{Chiang
  et~al.}{2021}]{Chiang2021}
Chiang B.~T.,  Schive H.~Y.,   Chiueh T.,  2021, \mn@doi [Physical Review D]
  {10.1103/PHYSREVD.103.103019/FIGURES/19/MEDIUM}, 103, 103019

\bibitem[\protect\citeauthoryear{Chomaz, Ferrier-Barbut, Ferlaino,
  Laburthe-Tolra, Lev  \& Pfau}{Chomaz et~al.}{2022}]{Chomaz_2023}
Chomaz L.,  Ferrier-Barbut I.,  Ferlaino F.,  Laburthe-Tolra B.,  Lev B.~L.,
  Pfau T.,  2022, \mn@doi [Reports on Progress in Physics]
  {10.1088/1361-6633/aca814}, 86, 026401

\bibitem[\protect\citeauthoryear{Cockburn, Negretti, Proukakis  \&
  Henkel}{Cockburn et~al.}{2011}]{Cockburn2011}
Cockburn S.~P.,  Negretti A.,  Proukakis N.,   Henkel C.,  2011, Physical
  Review A, 83, 043619

\bibitem[\protect\citeauthoryear{Dabo, Kozinsky, Singh-Miller  \& Marzari}{Dabo
  et~al.}{2008}]{Dabo2008}
Dabo I.,  Kozinsky B.,  Singh-Miller N.~E.,   Marzari N.,  2008, \mn@doi [Phys.
  Rev. B] {10.1103/PhysRevB.77.115139}, 77, 115139

\bibitem[\protect\citeauthoryear{Dawoodbhoy, Shapiro  \&
  Rindler-Daller}{Dawoodbhoy et~al.}{2021}]{Dawoodbhoy2021}
Dawoodbhoy T.,  Shapiro P.~R.,   Rindler-Daller T.,  2021, \mn@doi [Monthly
  Notices of the Royal Astronomical Society] {10.1093/MNRAS/STAB1859}, 506,
  2418

\bibitem[\protect\citeauthoryear{Delgado \& Mu{\~{n}}oz~Mateo}{Delgado \&
  Mu{\~{n}}oz~Mateo}{2022}]{Delgado2022}
Delgado V.,  Mu{\~{n}}oz~Mateo A.,  2022, \mn@doi [Monthly Notices of the Royal
  Astronomical Society] {10.1093/MNRAS/STAC3386}, 518, 4064

\bibitem[\protect\citeauthoryear{Desjacques, Kehagias  \& Riotto}{Desjacques
  et~al.}{2018}]{Desjacques:2017fmf}
Desjacques V.,  Kehagias A.,   Riotto A.,  2018, \mn@doi [Physical Review D]
  {10.1103/PhysRevD.97.023529}, 97, 023529

\bibitem[\protect\citeauthoryear{Dome, Fialkov, Mocz, Schäfer, Boylan-Kolchin
  \& Vogelsberger}{Dome et~al.}{2022}]{Dome2023}
Dome T.,  Fialkov A.,  Mocz P.,  Schäfer B.~M.,  Boylan-Kolchin M.,
  Vogelsberger M.,  2022, \mn@doi [Monthly Notices of the Royal Astronomical
  Society] {10.1093/mnras/stac3766}, 519, 4183

\bibitem[\protect\citeauthoryear{{Dutta Chowdhury}, van~den Bosch, Robles, van
  Dokkum, Schive, Chiueh  \& Broadhurst}{{Dutta Chowdhury}
  et~al.}{2021}]{DuttaChowdhury2021}
{Dutta Chowdhury} D.,  van~den Bosch F.~C.,  Robles V.~H.,  van Dokkum P.,
  Schive H.-Y.,  Chiueh T.,   Broadhurst T.,  2021, \mn@doi [The Astrophysical
  Journal] {10.3847/1538-4357/ac043f}, 916, 27

\bibitem[\protect\citeauthoryear{Fan}{Fan}{2016}]{Fan:2016rda}
Fan J.,  2016, \mn@doi [Phys. Dark Univ.] {10.1016/j.dark.2016.10.005}, 14, 84

\bibitem[\protect\citeauthoryear{{Ferreira}}{{Ferreira}}{2021}]{2021A&ARv..29....7F}
{Ferreira} E. G.~M.,  2021, \mn@doi [\aapr] {10.1007/s00159-021-00135-6}, \href
  {https://ui.adsabs.harvard.edu/abs/2021A&ARv..29....7F} {29, 7}

\bibitem[\protect\citeauthoryear{Foidl, Rindler-Daller  \& Zeilinger}{Foidl
  et~al.}{2023}]{Foidl2023}
Foidl H.,  Rindler-Daller T.,   Zeilinger W.~W.,  2023, \mn@doi [Phys. Rev. D]
  {10.1103/PhysRevD.108.043012}, 108, 043012

\bibitem[\protect\citeauthoryear{Galazo~Garc\'{\i}a, Brax  \&
  Valageas}{Galazo~Garc\'{\i}a et~al.}{2024}]{Valageas2024}
Galazo~Garc\'{\i}a R.,  Brax P.,   Valageas P.,  2024, \mn@doi [Phys. Rev. D]
  {10.1103/PhysRevD.109.043516}, 109, 043516

\bibitem[\protect\citeauthoryear{Garrett, Wright  \& Davis}{Garrett
  et~al.}{2013}]{PhysRevA.87.063611}
Garrett M.~C.,  Wright T.~M.,   Davis M.~J.,  2013, \mn@doi [Phys. Rev. A]
  {10.1103/PhysRevA.87.063611}, 87, 063611

\bibitem[\protect\citeauthoryear{Glennon \& Prescod-Weinstein}{Glennon \&
  Prescod-Weinstein}{2021}]{Glennon2020-2}
Glennon N.,  Prescod-Weinstein C.,  2021, \mn@doi [Physical Review D]
  {10.1103/PHYSREVD.104.083532/FIGURES/12/MEDIUM}, 104, 083532

\bibitem[\protect\citeauthoryear{Glyde, Azuah  \& Stirling}{Glyde
  et~al.}{2000}]{helium-depletion}
Glyde H.~R.,  Azuah R.~T.,   Stirling W.~G.,  2000, \mn@doi [Phys. Rev. B]
  {10.1103/PhysRevB.62.14337}, 62, 14337

\bibitem[\protect\citeauthoryear{Goodman}{Goodman}{2000}]{Goodman:2000tg}
Goodman J.,  2000, \mn@doi [New Astron.] {10.1016/S1384-1076(00)00015-4}, 5,
  103

\bibitem[\protect\citeauthoryear{Guzman \& Urena-Lopez}{Guzman \&
  Urena-Lopez}{2006}]{Guzman:2006yc}
Guzman F.~S.,  Urena-Lopez L.~A.,  2006, \mn@doi [Astrophys. J.]
  {10.1086/504508}, 645, 814

\bibitem[\protect\citeauthoryear{Harko}{Harko}{2011}]{Harko2011zt}
Harko T.,  2011, \mn@doi [Phys. Rev. D] {10.1103/PhysRevD.83.123515}, 83,
  123515

\bibitem[\protect\citeauthoryear{Harko \& Madarassy}{Harko \&
  Madarassy}{2012}]{Harko2012}
Harko T.,  Madarassy E.~J.,  2012, \mn@doi [Journal of Cosmology and
  Astroparticle Physics] {10.1088/1475-7516/2012/01/020}, 2012, 020

\bibitem[\protect\citeauthoryear{Hartman, Winther  \& Mota}{Hartman
  et~al.}{2022}]{Hartman2022}
Hartman S.~T.,  Winther H.~A.,   Mota D.~F.,  2022, \mn@doi [Journal of
  Cosmology and Astroparticle Physics] {10.1088/1475-7516/2022/02/005}, 2022

\bibitem[\protect\citeauthoryear{Hu, Barkana  \& Gruzinov}{Hu
  et~al.}{2000}]{Hu2000}
Hu W.,  Barkana R.,   Gruzinov A.,  2000, \mn@doi [Physical Review Letters]
  {10.1103/PhysRevLett.85.1158}, 85, 1158

\bibitem[\protect\citeauthoryear{Huang}{Huang}{2000}]{huang2000statistical}
Huang K.,  2000, Statistical Mechanics.
John Wiley and Sons

\bibitem[\protect\citeauthoryear{Hui}{Hui}{2021}]{Hui2021}
Hui L.,  2021, \mn@doi [Annual Review of Astronomy and Astrophysics]
  {10.1146/ANNUREV-ASTRO-120920-010024}, 59, 247

\bibitem[\protect\citeauthoryear{{Hui}, {Joyce}, {Landry}  \& {Li}}{{Hui}
  et~al.}{2021}]{2021JCAP...01..011H}
{Hui} L.,  {Joyce} A.,  {Landry} M.~J.,   {Li} X.,  2021, \mn@doi [\jcap]
  {10.1088/1475-7516/2021/01/011}, \href
  {https://ui.adsabs.harvard.edu/abs/2021JCAP...01..011H} {2021, 011}

\bibitem[\protect\citeauthoryear{Indjin, Liu, Proukakis  \& Rigopoulos}{Indjin
  et~al.}{2024}]{Indjin2024}
Indjin M.,  Liu I.-K.,  Proukakis N.~P.,   Rigopoulos G.,  2024, \mn@doi
  [Physical Review D] {10.1103/PhysRevD.109.103518}, 109, 103518

\bibitem[\protect\citeauthoryear{Indjin, Liu, Proukakis  \& Rigopoulos}{Indjin
  et~al.}{2025}]{Indjin:2025lqs}
Indjin M.,  Liu I.-K.,  Proukakis N.~P.,   Rigopoulos G.,  2025,  \mn@doi [arXiv e-prints] {10.48550/arXiv.2502.04838}, \href
  {https://arxiv.org/abs/2502.04838} {arXiv:2502.04838}


\bibitem[\protect\citeauthoryear{Khelashvili, Rudakovskyi  \&
  Hossenfelder}{Khelashvili et~al.}{2023}]{Khelashvili2023}
Khelashvili M.,  Rudakovskyi A.,   Hossenfelder S.,  2023, \mn@doi [Monthly
  Notices of the Royal Astronomical Society] {10.1093/mnras/stad1595}, 523,
  3393

\bibitem[\protect\citeauthoryear{{Khlopov}, {Malomed}  \&
  {Zeldovich}}{{Khlopov} et~al.}{1985}]{1985MNRAS.215..575K}
{Khlopov} M.~I.,  {Malomed} B.~A.,   {Zeldovich} I.~B.,  1985, \mn@doi [\mnras]
  {10.1093/mnras/215.4.575}, \href
  {https://ui.adsabs.harvard.edu/abs/1985MNRAS.215..575K} {215, 575}

\bibitem[\protect\citeauthoryear{Kirkpatrick, Mirasola  \&
  Prescod-Weinstein}{Kirkpatrick et~al.}{2020}]{chanda_formation_early}
Kirkpatrick K.,  Mirasola A.~E.,   Prescod-Weinstein C.,  2020, \mn@doi [Phys.
  Rev. D] {10.1103/PhysRevD.102.103012}, 102, 103012

\bibitem[\protect\citeauthoryear{Kirkpatrick, Mirasola  \&
  Prescod-Weinstein}{Kirkpatrick et~al.}{2022}]{chanda_formation}
Kirkpatrick K.,  Mirasola A.~E.,   Prescod-Weinstein C.,  2022, \mn@doi [Phys.
  Rev. D] {10.1103/PhysRevD.106.043512}, 106, 043512

\bibitem[\protect\citeauthoryear{{Kobayashi} \& {Tsubota}}{{Kobayashi} \&
  {Tsubota}}{2005}]{2005JPSJ...74.3248K}
{Kobayashi} M.,  {Tsubota} M.,  2005, \mn@doi [Journal of the Physical Society
  of Japan] {10.1143/JPSJ.74.3248}, \href
  {https://ui.adsabs.harvard.edu/abs/2005JPSJ...74.3248K} {74, 3248}

\bibitem[\protect\citeauthoryear{Korshynska, Bidasyuk, Gorbar, Jia  \&
  Yakimenko}{Korshynska et~al.}{2023}]{Korshynska_2023}
Korshynska K.,  Bidasyuk Y.~M.,  Gorbar E.~V.,  Jia J.,   Yakimenko A.~I.,
  2023, \mn@doi [The European Physical Journal C]
  {10.1140/epjc/s10052-023-11548-1}, 83

\bibitem[\protect\citeauthoryear{Korshynska, Prykhodko, Gorbar, Jia  \&
  Yakimenko}{Korshynska et~al.}{2025}]{Korshynska_2025}
Korshynska K.,  Prykhodko O.,  Gorbar E.,  Jia J.,   Yakimenko A.,  2025,
  \mn@doi [Physical Review D] {10.1103/physrevd.111.023006}, 111

\bibitem[\protect\citeauthoryear{Laroche, Gilman, Li, Bovy  \& Du}{Laroche
  et~al.}{2022}]{Laroche2022-lensing}
Laroche A.,  Gilman D.,  Li X.,  Bovy J.,   Du X.,  2022, \mn@doi [Monthly
  Notices of the Royal Astronomical Society] {10.1093/mnras/stac2677}, 517,
  1867

\bibitem[\protect\citeauthoryear{{Leadbeater}, {Winiecki}, {Samuels},
  {Barenghi}  \& {Adams}}{{Leadbeater} et~al.}{2001}]{Leadbeater2001}
{Leadbeater} M.,  {Winiecki} T.,  {Samuels} D.~C.,  {Barenghi} C.~F.,   {Adams}
  C.~S.,  2001, \mn@doi [\prl] {10.1103/PhysRevLett.86.1410}, \href
  {https://ui.adsabs.harvard.edu/abs/2001PhRvL..86.1410L} {86, 1410}

\bibitem[\protect\citeauthoryear{{Leadbeater}, {Samuels}, {Barenghi}  \&
  {Adams}}{{Leadbeater} et~al.}{2002}]{Leadbeater2002}
{Leadbeater} M.,  {Samuels} D.~C.,  {Barenghi} C.~F.,   {Adams} C.~S.,  2002,
  \mn@doi [arXiv e-prints] {10.48550/arXiv.cond-mat/0205588}, \href
  {https://ui.adsabs.harvard.edu/abs/2002cond.mat..5588L} {pp
  cond--mat/0205588}

\bibitem[\protect\citeauthoryear{{Leadbeater}, {Samuels}, {Barenghi}  \&
  {Adams}}{{Leadbeater} et~al.}{2003}]{Leadbeater2003}
{Leadbeater} M.,  {Samuels} D.~C.,  {Barenghi} C.~F.,   {Adams} C.~S.,  2003,
  \mn@doi [\pra] {10.1103/PhysRevA.67.015601}, \href
  {https://ui.adsabs.harvard.edu/abs/2003PhRvA..67a5601L} {67, 015601}

\bibitem[\protect\citeauthoryear{Lee \& Koh}{Lee \& Koh}{1996}]{Lee:1995af}
Lee J.-w.,  Koh I.-g.,  1996, \mn@doi [Phys. Rev. D]
  {10.1103/PhysRevD.53.2236}, 53, 2236

\bibitem[\protect\citeauthoryear{Leggett}{Leggett}{2006}]{leggett2006quantum}
Leggett A.,  2006, Quantum Liquids: Bose Condensation and Cooper Pairing in
  Condensed-matter Systems.
Oxford Graduate Texts, OUP Oxford, \url
  {https://books.google.co.uk/books?id=kywSDAAAQBAJ}

\bibitem[\protect\citeauthoryear{Lelli, McGaugh  \& Schombert}{Lelli
  et~al.}{2016}]{SPARC}
Lelli F.,  McGaugh S.~S.,   Schombert J.~M.,  2016, \mn@doi [The Astronomical
  Journal] {10.3847/0004-6256/152/6/157}, 152, 157

\bibitem[\protect\citeauthoryear{Levkov, Panin  \& Tkachev}{Levkov
  et~al.}{2018}]{Levkov2018}
Levkov D.~G.,  Panin A.~G.,   Tkachev I.~I.,  2018, \mn@doi [Phys. Rev. Lett.]
  {10.1103/PhysRevLett.121.151301}, 121, 151301

\bibitem[\protect\citeauthoryear{Li, Rindler-Daller  \& Shapiro}{Li
  et~al.}{2014}]{Li2014}
Li B.,  Rindler-Daller T.,   Shapiro P.~R.,  2014, \mn@doi [Physical Review D -
  Particles, Fields, Gravitation and Cosmology]
  {10.1103/PHYSREVD.89.083536/FIGURES/6/MEDIUM}, 89, 083536

\bibitem[\protect\citeauthoryear{Li, Shapiro  \& Rindler-Daller}{Li
  et~al.}{2017}]{Li2017}
Li B.,  Shapiro P.~R.,   Rindler-Daller T.,  2017, \mn@doi [Physical Review D]
  {10.1103/PHYSREVD.96.063505/FIGURES/13/MEDIUM}, 96, 063505

\bibitem[\protect\citeauthoryear{Li, Hui  \& Yavetz}{Li et~al.}{2020}]{Li2020}
Li X.,  Hui L.,   Yavetz T.~D.,  2020, \mn@doi [Physical
  Review D] {10.1103/PhysRevD.103.023508}, 103, 023508

\bibitem[\protect\citeauthoryear{Liao, Su, Schive, Kunkel, Huang  \&
  Chiueh}{Liao et~al.}{2025}]{schive2025}
Liao P.-Y.,  Su G.-M.,  Schive H.-Y.,  Kunkel A.,  Huang H.,   Chiueh T.,
  2025, Deciphering the Soliton-Halo Relation in Fuzzy Dark Matter (\mn@eprint
  {arXiv} {2412.09908}), \url {https://arxiv.org/abs/2412.09908}

\bibitem[\protect\citeauthoryear{Lin, Schive, Wong  \& Chiueh}{Lin
  et~al.}{2018}]{Lin2018}
Lin S.-C.,  Schive H.-Y.,  Wong S.-K.,   Chiueh T.,  2018, \mn@doi [Physical
  Review D] {10.1103/PhysRevD.97.103523}, 97, 103523



\bibitem[\protect\citeauthoryear{Liu, Proukakis  \& Rigopoulos}{Liu
  et~al.}{2023}]{Liu2023}
Liu I.-K.,  Proukakis N.~P.,   Rigopoulos G.,  2023, \mn@doi [Monthly Notices
  of the Royal Astronomical Society] {10.1093/mnras/stad591}, 521, 3625

\bibitem[\protect\citeauthoryear{Lopes, Eigen, Navon, Clément, Smith  \&
  Hadzibabic}{Lopes et~al.}{2017}]{Lopes_2017}
Lopes R.,  Eigen C.,  Navon N.,  Clément D.,  Smith R.~P.,   Hadzibabic Z.,
  2017, \mn@doi [Physical Review Letters] {10.1103/physrevlett.119.190404}, 119

\bibitem[\protect\citeauthoryear{Magaña \& Matos}{Magaña \&
  Matos}{2012}]{Magana2012}
Magaña J.,  Matos T.,  2012, \mn@doi [Journal of Physics: Conference Series]
  {10.1088/1742-6596/378/1/012012}, 378, 012012

\bibitem[\protect\citeauthoryear{{Marsh}}{{Marsh}}{2016}]{2016PhR...643....1M}
{Marsh} D. J.~E.,  2016, \mn@doi [Physics Reports]
  {10.1016/j.physrep.2016.06.005}, \href
  {https://ui.adsabs.harvard.edu/abs/2016PhR...643....1M} {643, 1}

\bibitem[\protect\citeauthoryear{Marsh \& Niemeyer}{Marsh \&
  Niemeyer}{2019}]{Marsh2019}
Marsh D.~J.,  Niemeyer J.~C.,  2019, \mn@doi [Physical Review Letters]
  {10.1103/PhysRevLett.123.051103}, 123, 051103

\bibitem[\protect\citeauthoryear{Marsh \& Pop}{Marsh \& Pop}{2015}]{Marsh2015}
Marsh D. J.~E.,  Pop A.-R.,  2015, \mn@doi [Monthly Notices of the Royal
  Astronomical Society] {10.1093/mnras/stv1050}, 451, 2479

\bibitem[\protect\citeauthoryear{Matos, Ureña-López  \& Lee}{Matos
  et~al.}{2024}]{Matos_Review_2024}
Matos T.,  Ureña-López L.~A.,   Lee J.-W.,  2024, \mn@doi [Frontiers in
  Astronomy and Space Sciences] {10.3389/fspas.2024.1347518}, 11

\bibitem[\protect\citeauthoryear{May \& Springel}{May \&
  Springel}{2021}]{May2021StructureDynamics}
May S.,  Springel V.,  2021, \mn@doi [Monthly Notices of the Royal Astronomical
  Society] {10.1093/MNRAS/STAB1764}, 506, 2603

\bibitem[\protect\citeauthoryear{Meinert \& Hofmann}{Meinert \&
  Hofmann}{2021}]{Meinert2021}
Meinert J.,  Hofmann R.,  2021, \mn@doi [Universe] {10.3390/universe7060198},
  7, 198

\bibitem[\protect\citeauthoryear{Membrado, Pacheco  \& Sa{\~n}udo}{Membrado
  et~al.}{1989}]{Membrado:1989bqo}
Membrado M.,  Pacheco A.~F.,   Sa{\~n}udo J.,  1989, \mn@doi [Phys. Rev. A]
  {10.1103/PhysRevA.39.4207}, 39, 4207

\bibitem[\protect\citeauthoryear{Mocz, Vogelsberger, Robles, Zavala,
  Boylan-Kolchin, Fialkov  \& Hernquist}{Mocz et~al.}{2017a}]{Mocz2017}
Mocz P.,  Vogelsberger M.,  Robles V.~H.,  Zavala J.,  Boylan-Kolchin M.,
  Fialkov A.,   Hernquist L.,  2017a, \mn@doi [Monthly Notices of the Royal
  Astronomical Society] {10.1093/mnras/stx1887}, 471, 4559

\bibitem[\protect\citeauthoryear{{Mocz}, {Vogelsberger}, {Robles}, {Zavala},
  {Boylan-Kolchin}, {Fialkov}  \& {Hernquist}}{{Mocz}
  et~al.}{2017b}]{2017MNRAS.471.4559M}
{Mocz} P.,  {Vogelsberger} M.,  {Robles} V.~H.,  {Zavala} J.,  {Boylan-Kolchin}
  M.,  {Fialkov} A.,   {Hernquist} L.,  2017b, \mn@doi [\mnras]
  {10.1093/mnras/stx1887}, \href
  {https://ui.adsabs.harvard.edu/abs/2017MNRAS.471.4559M} {471, 4559}

\bibitem[\protect\citeauthoryear{Mocz et~al.,}{Mocz et~al.}{2019}]{Mocz_2019}
Mocz P.,  et~al., 2019, \mn@doi [Phys. Rev. Lett.]
  {10.1103/PhysRevLett.123.141301}, 123, 141301

\bibitem[\protect\citeauthoryear{Mocz et~al.,}{Mocz et~al.}{2023}]{Mocz2023}
Mocz P.,  et~al., 2023, \mn@doi [Monthly Notices of the Royal Astronomical
  Society] {10.1093/mnras/stad694}, 2615, 2608

\bibitem[\protect\citeauthoryear{Moss}{Moss}{2024}]{Moss:2024mkc}
Moss I.~G.,  2024,
  \mn@doi [arXiv e-prints] {10.48550/arXiv.2407.13243}, \href
  {https://arxiv.org/abs/2407.13243} {arXiv:2407.13243}


\bibitem[\protect\citeauthoryear{Naraschewski \& Glauber}{Naraschewski \&
  Glauber}{1999}]{PhysRevA.59.4595}
Naraschewski M.,  Glauber R.~J.,  1999, \mn@doi [Phys. Rev. A]
  {10.1103/PhysRevA.59.4595}, 59, 4595

\bibitem[\protect\citeauthoryear{{Nore}, {Abid}  \& {Brachet}}{{Nore}
  et~al.}{1997a}]{1997PhFl....9.2644N}
{Nore} C.,  {Abid} M.,   {Brachet} M.~E.,  1997a, \mn@doi [Physics of Fluids]
  {10.1063/1.869473}, \href
  {https://ui.adsabs.harvard.edu/abs/1997PhFl....9.2644N} {9, 2644}

\bibitem[\protect\citeauthoryear{{Nore}, {Abid}  \& {Brachet}}{{Nore}
  et~al.}{1997b}]{1997PhRvL..78.3896N}
{Nore} C.,  {Abid} M.,   {Brachet} M.~E.,  1997b, \mn@doi [\prl]
  {10.1103/PhysRevLett.78.3896}, \href
  {https://ui.adsabs.harvard.edu/abs/1997PhRvL..78.3896N} {78, 3896}

\bibitem[\protect\citeauthoryear{{Numasato}, {Tsubota}  \& {L'Vov}}{{Numasato}
  et~al.}{2010}]{2010PhRvA..81f3630N}
{Numasato} R.,  {Tsubota} M.,   {L'Vov} V.~S.,  2010, \mn@doi [\pra]
  {10.1103/PhysRevA.81.063630}, \href
  {https://ui.adsabs.harvard.edu/abs/2010PhRvA..81f3630N} {81, 063630}

\bibitem[\protect\citeauthoryear{O'Dell, Giovanazzi, Kurizki  \& Akulin}{O'Dell
  et~al.}{2000}]{ODell:1999ewx}
O'Dell D.,  Giovanazzi S.,  Kurizki G.,   Akulin V.~M.,  2000, \mn@doi [Phys.
  Rev. Lett.] {10.1103/PhysRevLett.84.5687}, 84, 5687

\bibitem[\protect\citeauthoryear{O'Hare}{O'Hare}{2024}]{OHare2024}
O'Hare C. A.~J.,  2024, \mn@doi [PoS] {10.22323/1.454.0040}, COSMICWISPers, 040

\bibitem[\protect\citeauthoryear{Painter, Boylan-Kolchin, Mocz  \&
  Vogelsberger}{Painter et~al.}{2024}]{Mocz2024}
Painter C.~A.,  Boylan-Kolchin M.,  Mocz P.,   Vogelsberger M.,  2024, \mn@doi
  [Monthly Notices of the Royal Astronomical Society] {10.1093/mnras/stae1912},
  533, 2454

\bibitem[\protect\citeauthoryear{Peebles}{Peebles}{2000}]{Peebles:2000yy}
Peebles P. J.~E.,  2000, \mn@doi [Astrophys. J. Lett.] {10.1086/312677}, 534,
  L127

\bibitem[\protect\citeauthoryear{{Penrose} \& {Onsager}}{{Penrose} \&
  {Onsager}}{1956}]{1956PhRv..104..576P}
{Penrose} O.,  {Onsager} L.,  1956, \mn@doi [Physical Review]
  {10.1103/PhysRev.104.576}, \href
  {https://ui.adsabs.harvard.edu/abs/1956PhRv..104..576P} {104, 576}

\bibitem[\protect\citeauthoryear{{Pethick} \& {Smith}}{{Pethick} \&
  {Smith}}{2008}]{2008bcdg.book.....P}
{Pethick} C.~J.,  {Smith} H.,  2008, {Bose-Einstein Condensation in Dilute
  Gases, Cambridge Press}

\bibitem[\protect\citeauthoryear{Pitaevskii \& Stringari}{Pitaevskii \&
  Stringari}{2003}]{Pitaevskii2003}
Pitaevskii L.~P.,  Stringari S.,  2003, {Bose-Einstein Condensation}.
Clarendon Press

\bibitem[\protect\citeauthoryear{Powell, Vegetti, McKean, White, Ferreira, May
  \& Spingola}{Powell et~al.}{2023}]{Powell2023-lensing}
Powell D.~M.,  Vegetti S.,  McKean J.~P.,  White S. D.~M.,  Ferreira E. G.~M.,
  May S.,   Spingola C.,  2023, \mn@doi [Monthly Notices of the Royal
  Astronomical Society: Letters] {10.1093/mnrasl/slad074}, 524, L84

\bibitem[\protect\citeauthoryear{Proukakis}{Proukakis}{2023}]{proukakis2023encyclopedia}
Proukakis N.~P.,  2023, in Chakraborty T.,  ed., , Encyclopedia of Condensed
  matter Physics, 2nd Edition.
Elsevier (\mn@eprint {arXiv} {2304.09541}), \url
  {https://arxiv.org/abs/2304.09541}

\bibitem[\protect\citeauthoryear{Proukakis, Rigopoulos  \& Soto}{Proukakis
  et~al.}{2023}]{Proukakis:2023nmm}
Proukakis N.~P.,  Rigopoulos G.,   Soto A.,  2023, \mn@doi [Physical Review D]
  {10.1103/PhysRevD.108.083513}, 108, 083513

\bibitem[\protect\citeauthoryear{Proukakis, Rigopoulos  \& Soto}{Proukakis
  et~al.}{2024}]{Proukakis:2023txk}
Proukakis N.~P.,  Rigopoulos G.,   Soto A.,  2024, \mn@doi [Physical Review D]
  {10.1103/PhysRevD.110.023504}, 110, 023504

\bibitem[\protect\citeauthoryear{Proukakis, Rigopoulos  \& Soto}{Proukakis
  et~al.}{2025a}]{proukakis2025dipolar}
Proukakis N.~P.,  Rigopoulos G.,   Soto A.,  2025a, Dynamical Dipolar
  Condensate Finite Temperature Stochastic Gross--Pitaevskii--Boltzmann Model
  (\mn@eprint {arXiv} {2407.20178}), \url {https://arxiv.org/abs/2407.20178}

\bibitem[\protect\citeauthoryear{Proukakis, Rigopoulos  \& Soto}{Proukakis
  et~al.}{2025b}]{Proukakis:PhysRevD.111.023505}
Proukakis N.~P.,  Rigopoulos G.,   Soto A.,  2025b, \mn@doi [Phys. Rev. D]
  {10.1103/PhysRevD.111.023505}, 111, 023505

\bibitem[\protect\citeauthoryear{Rindler-Daller}{Rindler-Daller}{2023}]{Rindler-Daller:2022fjt}
Rindler-Daller T.,  2023, \mn@doi [Front. Astron. Space Sci.]
  {10.3389/fspas.2023.1121920}, 10, 1121920

\bibitem[\protect\citeauthoryear{Rindler-Daller \& Shapiro}{Rindler-Daller \&
  Shapiro}{2012}]{Rindler-Daller:2011afd}
Rindler-Daller T.,  Shapiro P.~R.,  2012, \mn@doi [Monthly Notices of the Royal
  Astronomical Society] {10.1111/j.1365-2966.2012.20588.x}, 422, 135

\bibitem[\protect\citeauthoryear{Rindler-Daller \& Shapiro}{Rindler-Daller \&
  Shapiro}{2014}]{Rindler-Daller2014}
Rindler-Daller T.,  Shapiro P.~R.,  2014, \mn@doi [Modern Physics Letters A]
  {10.1142/S021773231430002X}, 29

\bibitem[\protect\citeauthoryear{Salasnich \& Yakimenko}{Salasnich \&
  Yakimenko}{2025}]{Salasnich2025}
Salasnich L.,  Yakimenko A.,  2025, \mn@doi [Phys. Dark Univ.]
  {10.1016/j.dark.2025.101973}, 49, 101973

\bibitem[\protect\citeauthoryear{Schive, Chiueh  \& Broadhurst}{Schive
  et~al.}{2014a}]{Schive2014}
Schive H.~Y.,  Chiueh T.,   Broadhurst T.,  2014a, \mn@doi [Nature Physics]
  {10.1038/nphys2996}, 10, 496

\bibitem[\protect\citeauthoryear{Schive, Liao, Woo, Wong, Chiueh, Broadhurst
  \& Hwang}{Schive et~al.}{2014b}]{Schive2014a}
Schive H.~Y.,  Liao M.~H.,  Woo T.~P.,  Wong S.~K.,  Chiueh T.,  Broadhurst T.,
    Hwang W.~Y.,  2014b, \mn@doi [Physical Review Letters]
  {10.1103/PhysRevLett.113.261302}, 113, 1

\bibitem[\protect\citeauthoryear{{Schive}, {Chiueh}  \& {Broadhurst}}{{Schive}
  et~al.}{2020}]{Schive2020PhRvL.124t1301S}
{Schive} H.-Y.,  {Chiueh} T.,   {Broadhurst} T.,  2020, \mn@doi [\prl]
  {10.1103/PhysRevLett.124.201301}, \href
  {https://ui.adsabs.harvard.edu/abs/2020PhRvL.124t1301S} {124, 201301}

\bibitem[\protect\citeauthoryear{Schobesberger, Rindler-Daller  \&
  Shapiro}{Schobesberger et~al.}{2021}]{Schobesberger:2021ghi}
Schobesberger S.~O.,  Rindler-Daller T.,   Shapiro P.~R.,  2021, \mn@doi [Mon.
  Not. Roy. Astron. Soc.] {10.1093/mnras/stab1153}, 505, 802

\bibitem[\protect\citeauthoryear{Schwabe, Niemeyer  \& Engels}{Schwabe
  et~al.}{2016}]{Schwabe2016}
Schwabe B.,  Niemeyer J.~C.,   Engels J.~F.,  2016, \mn@doi [Physical Review D]
  {10.1103/PhysRevD.94.043513}, 94, 1

\bibitem[\protect\citeauthoryear{Shapiro, Dawoodbhoy  \&
  Rindler-Daller}{Shapiro et~al.}{2021}]{Shapiro2021}
Shapiro P.~R.,  Dawoodbhoy T.,   Rindler-Daller T.,  2021, \mn@doi [Monthly
  Notices of the Royal Astronomical Society] {10.1093/MNRAS/STAB2884}, 509, 145

\bibitem[\protect\citeauthoryear{Sin}{Sin}{1994}]{PhysRevD.50.3650}
Sin S.-J.,  1994, \mn@doi [Phys. Rev. D] {10.1103/PhysRevD.50.3650}, 50, 3650

\bibitem[\protect\citeauthoryear{Sivakumar, Mishra, Hujeirat  \&
  Muruganandam}{Sivakumar et~al.}{2025}]{Sivakumar2025}
Sivakumar A.,  Mishra P.~K.,  Hujeirat A.~A.,   Muruganandam P.,  2025, \mn@doi
  [Phys. Rev. D] {10.1103/PhysRevD.111.083511}, 111, 083511

\bibitem[\protect\citeauthoryear{Stallovits \& Rindler-Daller}{Stallovits \&
  Rindler-Daller}{2025}]{RindlerDaller2025}
Stallovits M.,  Rindler-Daller T.,  2025, \mn@doi [Phys. Rev. D]
  {10.1103/PhysRevD.111.023046}, 111, 023046

\bibitem[\protect\citeauthoryear{{Tsubota} \& {Kasamatsu}}{{Tsubota} \&
  {Kasamatsu}}{2012}]{2012arXiv1202.1863T}
{Tsubota} M.,  {Kasamatsu} K.,  2012, arXiv e-prints, \href
  {https://ui.adsabs.harvard.edu/abs/2012arXiv1202.1863T} {p. arXiv:1202.1863}

\bibitem[\protect\citeauthoryear{Vegetti et~al.,}{Vegetti
  et~al.}{2024}]{Vegetti2024-lensing}
Vegetti S.,  et~al., 2024, \mn@doi [Space Sci. Rev.]
  {10.1007/s11214-024-01087-w}, 220, 58

\bibitem[\protect\citeauthoryear{Veltmaat, Niemeyer  \& Schwabe}{Veltmaat
  et~al.}{2018}]{Veltmaat2018}
Veltmaat J.,  Niemeyer J.~C.,   Schwabe B.,  2018, \mn@doi [Phys. Rev. D]
  {10.1103/PhysRevD.98.043509}, 98, 043509

\bibitem[\protect\citeauthoryear{Woo \& Chiueh}{Woo \& Chiueh}{2009}]{Woo_2009}
Woo T.-P.,  Chiueh T.,  2009, \mn@doi [The Astrophysical Journal]
  {10.1088/0004-637X/697/1/850}, 697, 850

\bibitem[\protect\citeauthoryear{Yavetz, Li  \& Hui}{Yavetz
  et~al.}{2022}]{Yavetz2021}
Yavetz T.~D.,  Li X.,   Hui L.,  2022, \mn@doi [Phys. Rev. D]
  {10.1103/PhysRevD.105.023512}, 105, 023512

\bibitem[\protect\citeauthoryear{Zagorac, Sands, Padmanabhan  \&
  Easther}{Zagorac et~al.}{2022}]{Zagorac2022}
Zagorac J.~L.,  Sands I.,  Padmanabhan N.,   Easther R.,  2022, \mn@doi [Phys.
  Rev. D] {10.1103/PhysRevD.105.103506}, 105, 103506

\bibitem[\protect\citeauthoryear{Zagorac, Kendall, Padmanabhan  \&
  Easther}{Zagorac et~al.}{2023}]{Zagorac2023}
Zagorac J.~L.,  Kendall E.,  Padmanabhan N.,   Easther R.,  2023, \mn@doi
  [Phys. Rev. D] {10.1103/PhysRevD.107.083513}, 107, 083513

\bibitem[\protect\citeauthoryear{Zimmermann, Alvey, Marsh, Fairbairn  \&
  Read}{Zimmermann et~al.}{2025}]{Zimmermann2025}
Zimmermann T.,  Alvey J.,  Marsh D. J.~E.,  Fairbairn M.,   Read J.~I.,  2025,
  \mn@doi [Phys. Rev. Lett.] {10.1103/PhysRevLett.134.151001}, 134, 151001

\bibitem[\protect\citeauthoryear{{Zuccher}, {Caliari}, {Baggaley}  \&
  {Barenghi}}{{Zuccher} et~al.}{2012}]{Zuccher2012}
{Zuccher} S.,  {Caliari} M.,  {Baggaley} A.~W.,   {Barenghi} C.~F.,  2012,
  \mn@doi [Physics of Fluids] {10.1063/1.4772198}, \href
  {https://ui.adsabs.harvard.edu/abs/2012PhFl...24l5108Z} {24, 125108}

\makeatother
\end{thebibliography}

\bsp	
\label{lastpage}
\appendix

\section{Shape parameters}
\label{app:shape-params}
Our previously-studied~\cite{Indjin2024} extended data simulation set 
of true ground state solutions of the GPPE [Eqs.~\eqref{eq:GPE1} and \eqref{eq:Poisson_0} in the main text] with $\Gamma_g \in (0,\, 10^5)$,  can be conveniently re-analyzed by a SG profile [Eq.~\eqref{eq:empirical} in main text].
For a fixed $M$, $m$ and $g$ (which fixes $\Gamma_g$), we self-consistently obtain the value of $\vartheta(\Gamma_g)$ matching the numerical profile.
This yields equation \eqref{eq:shape_fit} as an empirical relationship for the SG exponent. Given $\vartheta(\Gamma_g)$, the shape parameters for the SG profile, Eq.~\eqref{eq:superGauss}, can be analytically expressed as:    
\be
\eta_\text{SG}(\Gamma_g) = \left(\frac{1}{\ln 2}\right)^{3/\vartheta(\Gamma_g)}\frac{1}{\vartheta(\Gamma_g)} \, \Gamma\left(\frac{3}{\vartheta(\Gamma_g)}\right),
\ee

\be
\sigma_\text{SG}(\Gamma_g) = \frac{\vartheta(\Gamma_g)^2 (\ln 2)^{2/\vartheta(\Gamma_g)} }{8} \frac{\Gamma\left(\frac{2\vartheta(\Gamma_g)+1}{\vartheta(\Gamma_g)}\right)}{\Gamma\left(\frac{3}{\vartheta(\Gamma_g)}\right)},
\ee

\be
\nu_\text{SG} = \frac{1}{2\eta^2 \vartheta(\Gamma_g)^2 (\ln 2)^{5/\vartheta(\Gamma_g)}} \Gamma\left(\frac{5}{\vartheta(\Gamma_g)}\right) B\left(\frac{1}{2};\frac{2}{\vartheta(\Gamma_g)};\frac{3}{\vartheta(\Gamma_g)}\right),
\ee

\be
\zeta_\text{SG}(\Gamma_g) = \frac{1}{\vartheta(\Gamma_g) (\ln 2)^{3/\vartheta(\Gamma_g)}} \left(\frac{1}{2}\right)^{3/\vartheta(\Gamma_g) -1} \Gamma\left(\frac{3}{\vartheta(\Gamma_g)}\right) \frac{1}{8\pi \eta_\text{SG}^2},
\ee
and
\be
\alpha_\text{SG}(\Gamma_g) = \frac{1}{\vartheta(\Gamma_g) \eta_\text{SG}(\Gamma_g) (\ln 2)^{5/\vartheta(\Gamma_g)}} \Gamma\left(\frac{5}{\vartheta(\Gamma_g)}\right),
\ee
where $\Gamma(z)$ denotes the Gamma function and $B(z,a,b)$ is the incomplete Beta function.

\section{Core mass} \label{sec:appendix-b}

To get a better handle of when the soliton merger for a given interaction strength reaches a quasi-steady state for conducting our analysis, we consider here the evolution of the soliton core mass
via
\begin{equation}
M_\mathrm{core}=4\pi\int_0^{r_c}drr^2\rho(r) \;.
\label{eq:PO-core}
\end{equation}
Such evolution is shown in Fig.~\ref{fig:appendix} for the 4 main $g^\prime$ values considered in the text.
This allows us to confidently identify the optimal time span for averaging, as highlighted by the green dotted `averaging period' bar. For $\Gamma_g=0$ the core settles relatively quickly to an (apparently constant, or perhaps very slowly increasing) core mass with relatively small fluctuations, and in order to suppress any oscillatory features, we use a sufficiently long averaging period of $3.5 \tau$. However, to accommodate for the fact that mergers with  $\Gamma_g \neq 0$ can lead to significantly fluctuating cores, requiring longer evolution times to reach a steady-state, we begin sampling the equally-extended time-averaging period for PO mode extraction at a later evolution time,
as indicated in Fig.~\ref{fig:appendix}. 
For each of these cases, the figure also shows the corresponding mass predicted by the PO mode, obtained from Eq.~\eqref{eq:PO-core} with $\rho=\rho_{PO}$ over such averaging period: For $g \neq 0$, this is found to be less than the total numerical mass of the soliton, showcasing the dynamical nature of the condensate depletion discussed in the main text.

\begin{figure*}
 \centering
	\includegraphics[width=2\columnwidth]{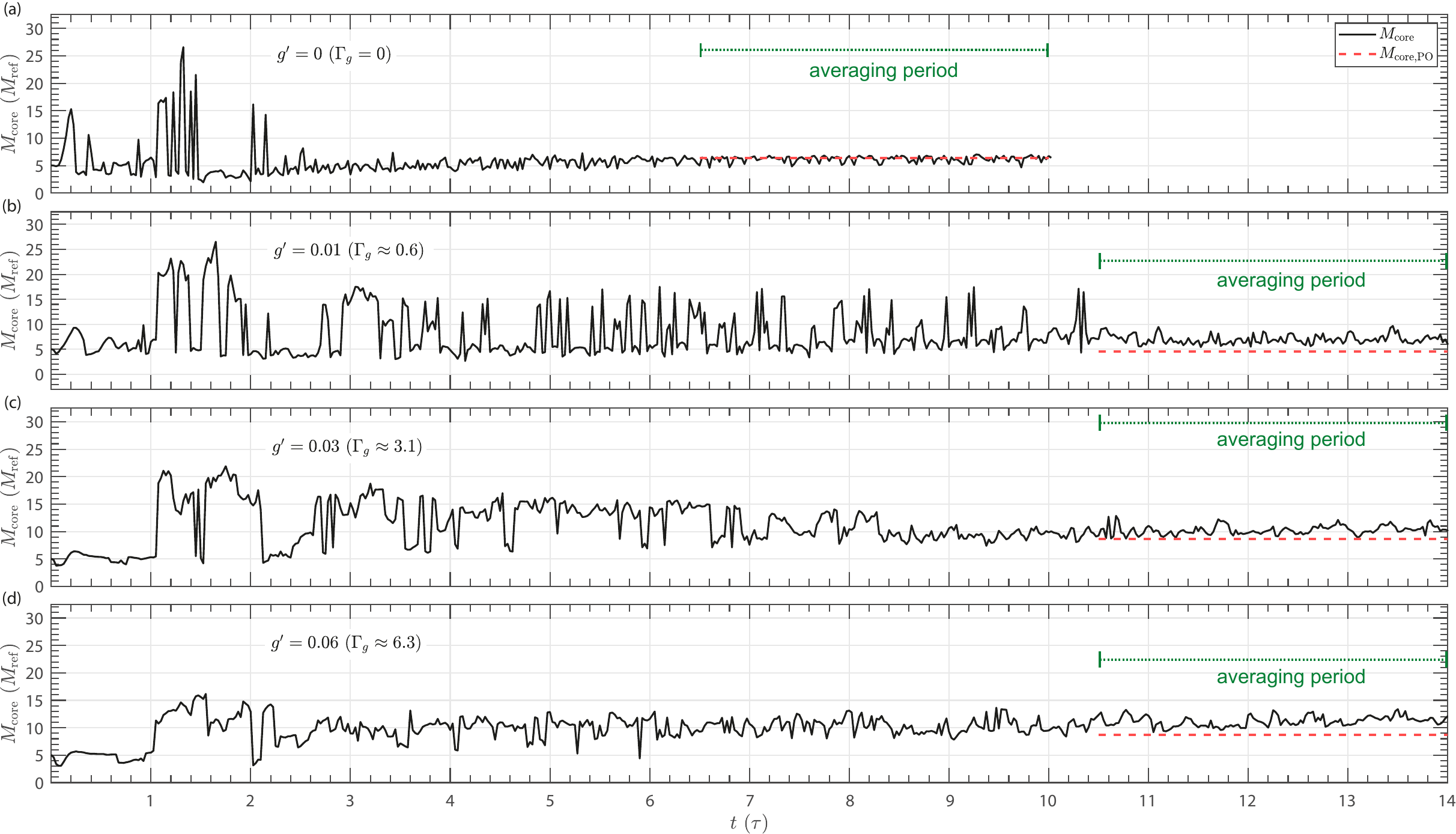}
    \caption{The core mass evolutions obtained from Eq.~\eqref{eq:PO-core} for (a) $g^\prime=0$, (b) $g^\prime=10^{-2}\approx0.3g_\ast$, (c) $g^\prime=3\times10^{-2}\approx g_\ast$ and (d) $g^\prime=0.06\approx6.3g_\ast$,
    with the averaging period of 3.5$\tau_{\rm ref}$ clearly highlighted in each case. The red dashed lines show the masses of the corresponding PO modes, which for $g' \neq 0$ are clearly less than the corresponding total masses within the same spatial region $r \leq r_c$.
    \label{fig:appendix}}
\end{figure*}
\end{document}